\documentclass[
twocolumn,
aps,
prl,
reprint,
superscriptaddress,
amsmath,amssymb
]{revtex4-1}
\usepackage[pdftex]{graphicx} % pdfLatex
\usepackage[hidelinks]{hyperref} % pdfLatex
\hypersetup{% hyperrefオプションリスト
 setpagesize=false,
 bookmarksnumbered=true,%
 bookmarksopen=true,%
 colorlinks=true,%
 linkcolor=blue,
 citecolor=blue,
}
\usepackage{braket}
\usepackage{times,multirow,amsfonts,bm,xspace,pifont}
\usepackage{soul}
\usepackage{ulem}
\usepackage{mathrsfs}
\begin{document}
\newcommand{\rr}{{\bm r}}
\newcommand{\q}{{\bm q}}
\renewcommand{\k}{{\bm k}}
\newcommand*\YY[1]{\textcolor{blue}{#1}}
\newcommand{\YYS}[1]{\textcolor{blue}{\sout{#1}}}
\newcommand*\YYWC[1]{\textcolor{green}{#1}}
\newcommand{\TK}[1]{{\color{red}{#1}}}
\newcommand*\TKS[1]{\textcolor{red}{\sout{#1}}}

\newcommand*\AD[1]{\textcolor{magenta}{#1}}
\newcommand{\ADS}[1]{\textcolor{magenta}{\sout{#1}}}

\newcommand*\BdG{{\rm BdG}}

% Title of paper
\title{
Spin-triplet superconductivity from quantum-geometry-induced ferromagnetic fluctuation
}

%authors
\author{Taisei Kitamura}
\email[]{kitamura.taisei.67m@st.kyoto-u.ac.jp}
\affiliation{Department of Physics, Graduate School of Science, Kyoto University, Kyoto 606-8502, Japan}

\author{Akito Daido}
%\email[]{}
%\affiliation{Department of Physics, Graduate School of Science, Kyoto University, Kyoto 606-8502, Japan}
\affiliation{Department of Physics, Graduate School of Science, Kyoto University, Kyoto 606-8502, Japan}

\author{Youichi Yanase}
%\email[]{yanase@scphys.kyoto-u.ac.jp}
\affiliation{Department of Physics, Graduate School of Science, Kyoto University, Kyoto 606-8502, Japan}

\date{\today}

\begin{abstract}
We show that quantum geometry induces ferromagnetic fluctuation resulting in spin-triplet superconductivity. The criterion for ferromagnetic fluctuation is clarified by analyzing contributions from the effective mass and quantum geometry. When the non-Kramers band degeneracy is present near the Fermi surface, the Fubini-Study quantum metric strongly favors ferromagnetic fluctuation. Solving the linearized gap equation with the effective interaction obtained by the random phase approximation, we show that the spin-triplet superconductivity is mediated by quantum-geometry-induced ferromagnetic fluctuation.
\end{abstract}

\maketitle

\textit{Introduction.---} 
Unconventional superconductivity beyond the canonical Bardeen-Cooper-Schrieffer theory shows rich physical phenomena including high-temperature superconductivity and topological superconductivity.
Various fluctuations arising from many-body interactions play the main role in the Cooper pairing for unconventional superconductivity, and low-dimensional fluctuations are particularly favorable. For example, it is argued that high-temperature superconductivity in cuprates is mediated by two-dimensional antiferromagnetic fluctuation~\cite{moriya2000spin,tsuei2000pairing,yanase2003theory}. Also, in iron-based high-temperature superconductors the extended $s$-wave pairing is mediated by orbital~\cite{kontani2010orbital,yanagi2010orbital,onari2012self} or antiferromagnetic~\cite{mazin2008unconventional,kuroki2008unconventinal} fluctuation~\cite{hosono2015iron, chubukov2012pairing,chubukov2016magnetism}.

However, searching for topological superconductivity~\cite{qi2011topological,alicea2012new,sato2016majorana,sato2017topological} with Majorana fermion~\cite{sato2009topological,sato2010topological,fu2010odd} is an unresolved problem of modern condensed matter physics, which is attributed to the fact that the platform for topological superconductivity is rare in nature. Spin-triplet superconductors are canonical candidates, and it is expected that ferromagnetic fluctuation mediates the spin-triplet Cooper pairing. However, candidate materials are restricted to a few heavy-fermion systems with three-dimensional multiple bands~\cite{sauls1994the,tou1998nonunitary,joynt2002the,saxena2000superconductivity,aoki2001coexistence,huy2007superconductivity,ran2019nearly,aoki2021unconventional}.

In the two-dimensional isotropic continuum models, ferromagnetic fluctuation is not favored because of the constant density of states (DOS), which may imply the absence of two-dimensional spin-triplet superconductivity.
Even for the anisotropic lattice systems, most quasi-two-dimensional superconductors do not show ferromagnetic fluctuation and antiferromagnetic fluctuations are rather ubiquitous, as we mentioned above for cuprates and iron-based compounds. 
Thus, spin-triplet superconductivity from ferromagnetic fluctuation is expected to require peculiar band structures, and the search for such systems is challenging for both materials and theoretical models.
In this Letter, nevertheless, we propose a guiding principle for realizing ferromagnetic fluctuation in two-dimensional systems by referring to the quantum geometry of Bloch electrons,  which is recently attracting much attention in various fields~\cite{marzari1997maximally,resta2011the,gao2014field,gao2019nonreciprocal,lapa2019semiclassical,daido2020thermodynami,kitamura2021thermodynamic,peotta2015superfluidity,liang2017band,torma2022superconductivity,rossi2021quantum,julku2021excitations,julku2021quantum,liao2021experimental,chen2023towards,iskin2023extracting,rhim2020quantum,watanabe2021chiral,ahn2020low,ahn2021riemannian}.

The importance of quantum geometry in superconductors has recently been recognized as it gives correction to the superfluid weight~\cite{peotta2015superfluidity,liang2017band,torma2022superconductivity,rossi2021quantum} 
.
In the flat-band systems ~\cite{julku2016geometric,julku2020superfluid,xie2020topology,hu2019geometric,tian2023evidence,torma2022superconductivity,rossi2021quantum}
the superfluid weight from Fermi-liquid theory vanishes, and the quantum geometric contribution determines the superfluid weight. 
The quantum geometry also plays essential roles in the monolayer FeSe~\cite{kitamura2022superconductivity} and some finite-momentum Cooper pairing states~\cite{kitamura2022quantum,chen2023pair,jiang2023pair,kitamura2022quantum-geometry,chazono2022piezoelectric}.
However, how quantum geometry affects the pairing mechanism of superconductivity has not been revealed. This work elucidates a way to create a pairing glue of unconventional superconductivity via quantum geometry.

To show that the quantum geometry enables strong ferromagnetic fluctuation in two-dimensional systems, resulting in spin-triplet superconductivity, we elucidate the criterion for ferromagnetic fluctuation
in the multi-band system with SU(2) symmetry.
We find that the criterion is given by the generalized electric susceptibility (GES) which is defined as a natural extension of the electric susceptibility to metals. The GES contains the terms obtained by the effective mass and the quantum geometry.

The key physics of quantum-geometry-induced ferromagnetic fluctuation, which is shown below, is nontrivial quantum geometry, especially Fubini-Study quantum metric~\cite{provost1980riemannian,resta2011the}, from non-Kramers band degeneracy.
As shown in this Letter, the dispersive Lieb lattice model with non-Kramers band degeneracy shows strong ferromagnetic fluctuation by this mechanism.
Solving the linearized gap equation with the effective interaction calculated by the random phase approximation (RPA), spin-triplet superconductivity is demonstrated.

\textit{Criterion for ferromagnetic fluctuation in multi-band Hubbard models.---}
\begin{figure}[tbp]
  \includegraphics[width=1.0\linewidth]{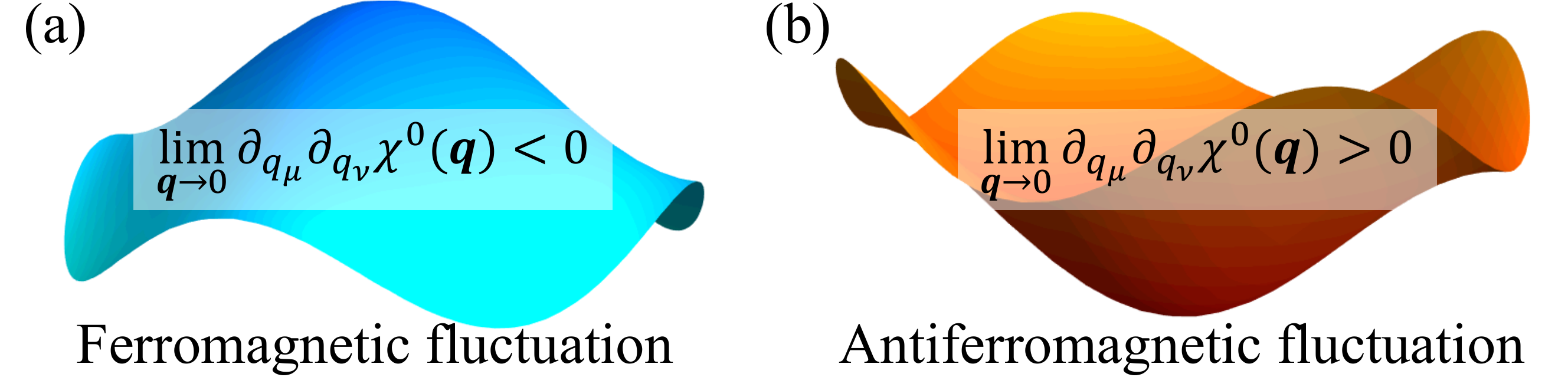}
  \centering
  \caption{Schematic figures for (a) ferromagnetic and (b) antiferromagnetic fluctuation.
  We illustrate the ${\bm q}$-dependence of $\chi^0(\bm q)$.
  \label{fig:schematic}}
\end{figure}
We consider the multi-band Hubbard model with SU(2) symmetry, which contains multiple degrees of freedom such as orbitals and sublattices~\cite{Supplemental}. The SU(2) symmetry means that the spin-orbit coupling and the magnetic field are absent. For the interacting Hamiltonian, we consider the onsite Coulomb interaction $U$ strong enough for the superconducting transition, by assuming strongly correlated materials. We then focus on the momentum dependence of the fluctuation, which mainly determines the superconducting symmetry~\cite{yanase2003theory}.
While we consider two-dimensional systems, the following discussions apply to three-dimensional systems.

Throughout this paper, $U$ is treated in the RPA scheme. When the system has only one band, the spin (charge) susceptibility $\chi_{\rm s(c)}(\bm q, i\Omega_n)$ can be obtained as $\chi_{\rm s(c)}(\bm q, i\Omega_n) = \chi_{\rm s(c)}^0(\bm q, i\Omega_n)/(1\mp \frac{U}{2}\chi_{\rm s(c)}^0(\bm q, i\Omega_n))$ by using the bare spin (charge) susceptibility of noninteracting systems, $\chi^0_{\rm s(c)}(\bm q, i\Omega_n)$. 
The interaction does not change the position of peaks in the momentum $\bm q$ space.
Therefore, also for most multi-band systems, it is expected that the momentum dependence of fluctuations arises from the bare susceptibility. 
Because the low-frequency spin (charge) fluctuation plays the dominant role in mediating superconductivity, hereafter we focus on the static fluctuations at $\Omega_n = 0$.

In multi-band systems with SU(2) symmetry, the bare spin/charge susceptibilities hold the relationship $\chi^0_{\rm s}(\bm q)=\chi^0_{\rm c}(\bm q)=2\chi^0(\bm q)$ with the bare susceptibility $\chi^0(\bm q)$. Thus, our main concern is the presence/absence of the peak of $\chi^0(\bm q)$ at $\bm q = 0$, corresponding to the presence/absence of ferromagnetic fluctuation. The structure of susceptibility $\chi^0(\bm q)$ around $\bm q = 0$ is determined by the curvature $\lim_{\bm q\rightarrow0}\partial_{q_\mu}\partial_{q_\nu}\chi^0(\bm q)$ with $\mu,\nu = x,y$~\cite{qchicomment}. As a result, the criterion for the ferromagnetic fluctuation is given by the sign of the curvature (see Fig.~\ref{fig:schematic}).
Ferromagnetic fluctuation may be present when $\lim_{\bm q\rightarrow0}\partial_{q_\mu}\partial_{q_\nu}\chi^0(\bm q)$ is negative. Otherwise, ferromagnetic fluctuation is prohibited.

The curvature $\lim_{\bm q\rightarrow0}\partial_{q_\mu}\partial_{q_\nu}\chi^0(\bm q)$ itself has a physical meaning. For the discussion, it is useful to consider the charge susceptibility in insulators at zero temperature, instead of the spin susceptibility. Based on the Kubo formula, the curvature expresses the correction to the charge density, $\delta\braket{\hat{n}(\bm r)}$, by the external electric field $E_\nu(\bm r)$ as,
$
\delta\braket{\hat{n}(\bm r)}=-\sum_{\mu\nu}\partial_{r_\nu}(\lim_{\bm q\rightarrow 0}\frac{1}{2}\partial_{q_\mu}\partial_{q_\nu}\chi_{\rm c}^0(\bm q)E_\nu(\bm r))
$
~\cite{Supplemental}.
%\TKS{where $\braket{\hat{n}(\bm r)}$ and $E(\bm r)$ are the charge density and the external electric field, respectively~\cite{Supplemental}. $\rho_0$ is the charge density in the absence of the external field.}
This means that the curvature $\lim_{\bm q=0}\partial_{q_\mu}\partial_{q_\nu}\chi^0(\bm q)$ is the electric susceptibility. Thus, by generalizing the concept of the electric susceptibility to metals, we define the generalized electric susceptibility (GES) as $\chi_{\rm e}^{0:\mu\nu} \equiv \lim_{\bm q=0}\partial_{q_\mu}\partial_{q_\nu}\chi^0(\bm q)$~\cite{Supplemental}.

\textit{Formula of GES.---} 
Here, we derive the formula of GES~\cite{Supplemental}, $\chi_{\rm e}^{0:\mu\nu} = \chi_{\rm e:geom}^{0:\mu\nu}+\chi_{\rm e:mass}^{0:\mu\nu}$, 
\begin{eqnarray}
%    \!&&\chi_{\rm e}^{0:\mu\nu} = \chi_{\rm e:geom}^{0:\mu\nu}+\chi_{\rm e:mass}^{0:\mu\nu}, \label{eq:1}\\
    &&\!\chi_{\rm e:geom}^{0:\mu\nu}=\notag\\
    &&\!2\sum_{n}\int \dfrac{d\bm k}{(2\pi)^2}\left(
    \dfrac{f^\prime(\epsilon_n(\bm k))}{2}g_n^{\mu\nu}(\bm k)+f(\epsilon_n(\bm k))X_n^{\mu\nu}(\bm k)
    \right),\notag\\\label{eq:2}\\
    &&\!\chi_{\rm e:mass}^{0:\mu\nu}=-2\sum_{n}\int \dfrac{d\bm k}{(2\pi)^2}
    \dfrac{f^{(2)}(\epsilon_n(\bm k))}{12}[m_n^{\mu\nu}(\bm k)]^{-1}
    , \label{eq:3}
\end{eqnarray}
where $\epsilon_n(\bm k)$ is the energy of the 
noninteracting Hamiltonian $\sigma_0\otimes H_0(\bm k)$, which follows $H_0(\bm k)\ket{u_n(\bm k)}=\epsilon_n(\bm k)\ket{u_n(\bm k)}$ with the Bloch wave function $\ket{u_n(\bm k)}$.
Note that $\sigma_0$ is the unit matrix of spin space and $n$ is the band index.
Thus, GES is given by the two terms, $\chi_{\rm e:geom}^{0:\mu\nu}$ and $\chi_{\rm e:mass}^{0:\mu\nu}$.
%\TKS{Equations~\eqref{eq:2}-\eqref{eq:3} show that the GES is given by the two terms, $\chi_{\rm e:geom}^{0:\mu\nu}$ and $\chi_{\rm e:mass}^{0:\mu\nu}$. }

The first term $\chi_{\rm e:geom}^{0:\mu\nu}$ named quantum geometric term is determined by the geometric quantities, namely, the Fubini-Study quantum metric 
$
g_n^{\mu\nu}(\bm k) = \sum_{m(\neq n)}A_{nm}^\mu(\bm k) A_{mn}^\nu(\bm k)+{\rm c.c.}
$
and the positional shift 
$
X_n^{\mu\nu}(\bm k) = \sum_{m(\neq n)}(A_{nm}^\mu(\bm k) A_{mn}^\nu(\bm k)+{\rm c.c.})/(\epsilon_m(\bm k)-\epsilon_n(\bm k))
$
with the Berry connection $A_{nm}^\mu(\bm k) = i\braket{\partial_{k_\mu}u_n(\bm k)\vert u_m(\bm k)}$. 
This term arises from purely interband effects and is absent in single-band systems. 
In this term, the contributions from the quantum metric and the positional shift are competitive. 
First, the quantum metric~\cite{provost1980riemannian,resta2011the}, which is the counterpart of the Berry curvature~\cite{berry1984quantal}, represents the distance between two adjacent states and is a positive definite tensor. Therefore, combined with negative $f^\prime(\epsilon_n(\bm k))$, the contribution from the quantum metric is always negative, favoring ferromagnetic fluctuation.
Second, the positional shift~\cite{gao2014field} means the shift of electrons by the external electric field. In insulators at zero temperature, the contribution from the positional shift corresponds to the well-known formula of electric susceptibility~\cite{aversa1995nonlinear}. This term can be rewritten as it is proportional to $F_{nm}(\bm k)(A_{nm}^\mu(\bm k) A_{mn}^\nu(\bm k)+{\rm c.c.})$ with the integrand of the Lindhard function, $F_{nm}(\bm k, \bm q) = (f(\epsilon_m(\bm k))-f(\epsilon_n(\bm k+\bm q)))/(\epsilon_n(\bm k+\bm q)-\epsilon_m(\bm k))\xrightarrow{\bm q\to 0} F_{nm}(\bm k)$. Therefore, this contribution is always positive, which favors antiferromagnetic fluctuation.

Importantly, both quantum metric and positional shift diverge at the non-Kramers band-degenerate point. Therefore, quantum geometry plays an essential role when non-Kramers band degeneracy exists.
However, the total geometric term does not diverge because of the cancellation of two contributions~\cite{Supplemental}.

The effective-mass term $\chi_{\rm e:mass}^{0:\mu\nu}$ of GES is the purely intraband effect and is determined by the band dispersion through the effective mass 
$
[m_n^{\mu\nu}(\bm k)]^{-1} = \partial_{k_\mu}\partial_{k_\nu}\epsilon_n(\bm k) 
$. 
In single-band systems, only this term is finite. This term can be positive and negative. 
For the hyperbolic dispersion $\epsilon_n(\bm k) = k^2/2m $,  the effective-mass term is zero because the DOS and effective mass are constants, which means the absence of ferromagnetic fluctuation~\cite{Supplemental}.

\textit{GES with non-Kramers band degeneracy.---} 
Because the non-Kramers band degeneracy enhances the quantum geometry, we focus on the Lieb lattice, which has been realized in ultracold atoms allowing us to tune the strength of $U$~\cite{gross2017quantum,taie2015coherent}, with the experimental test in mind. 
The Lieb lattice hosts the flat band with three-fold band degeneracy,
and the ground state shows the flat-band ferromagnetism~\cite{lieb1989two}.
To distinguish the quantum-geometry-induced ferromagnetic fluctuation from the flat-band ferromagnetism, 
we study the dispersive Lieb lattice model in which the second and third-nearest-neighbor hoppings are finite. Unlike the usual Lieb lattice with only the nearest-neighbor hopping, the flat band becomes dispersive and the three-fold band degeneracy at the $M$ point [${\bm k} =(\pi,\pi)$] is partially lifted, while the two-fold degeneracy remains protected by the $C_4$ rotation symmetry~\cite{Supplemental}.

\begin{figure}[tbp]
  \includegraphics[width=1.0\linewidth]{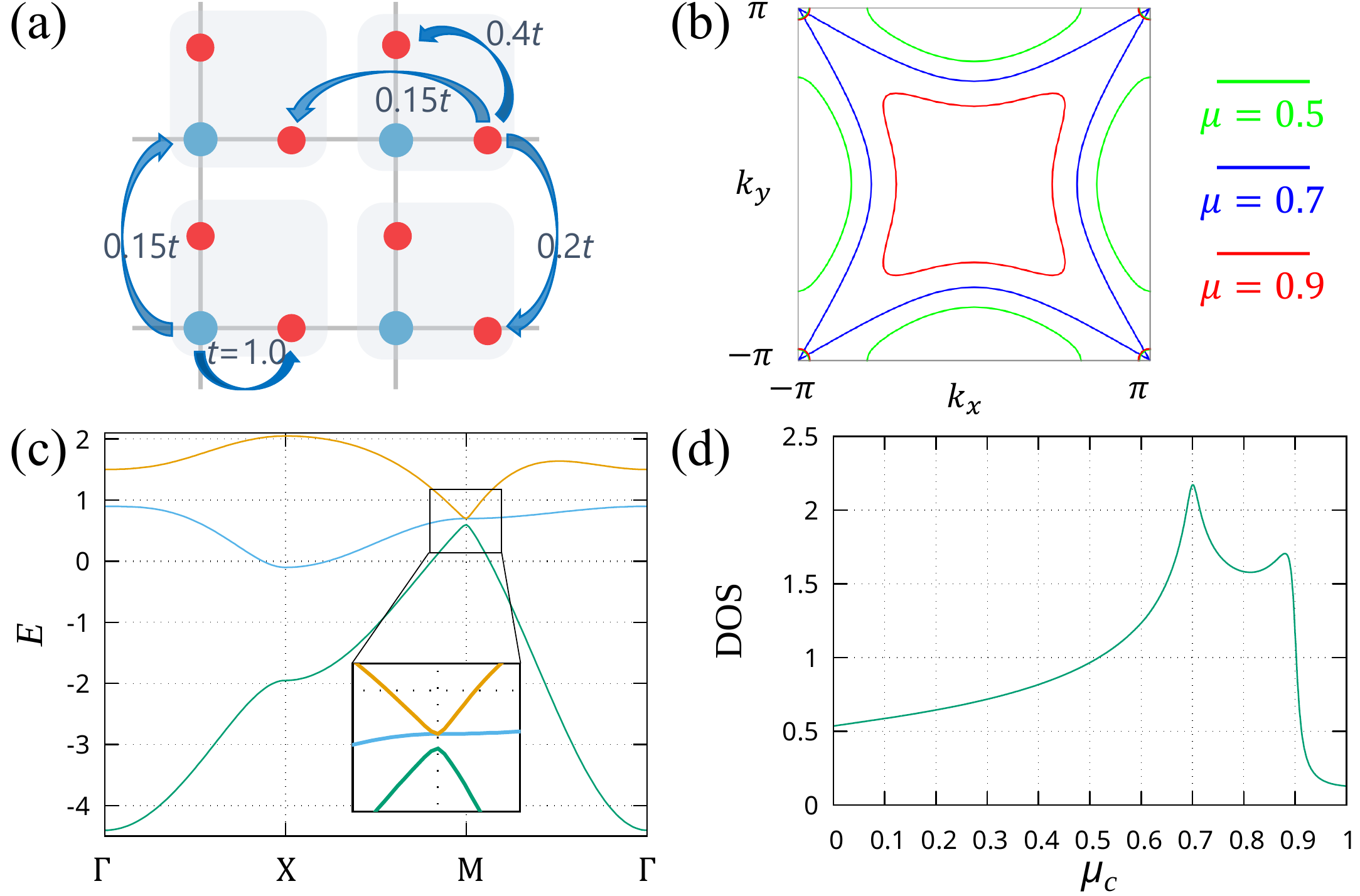}
  \centering
  \caption{(a) The dispersive Lieb lattice model, unit-cell (gray box), and hopping integrals (blue arrows). The first-nearest-neighbor hopping is taken as the unit of energy, $t=1$. (b) The Fermi surface for $\mu_{\rm c} = 0.5$ (green line), $0.7$ (blue line), and $0.9$ (red line). (c) The band dispersion. (d) The DOS. \label{fig:hk_Lieb_all}}
\end{figure}

The dispersive Lieb lattice model is illustrated in Fig.~\ref{fig:hk_Lieb_all}(a).
The Fermi surfaces for the chemical potential $\mu_{\rm c} = 0.5, 0.7$, and $0.9$ are shown in Fig.~\ref{fig:hk_Lieb_all}(b), and the band dispersion is in Fig.~\ref{fig:hk_Lieb_all}(c). The band-degenerate point lies on the Fermi surface, when $\mu_{\rm c} = 0.7$. 
As shown in Fig.~\ref{fig:hk_Lieb_all}(d), the maximum of DOS corresponds to $\mu_{\rm c} = 0.7$.

\begin{figure}[tbp]
  \includegraphics[width=1.0\linewidth]{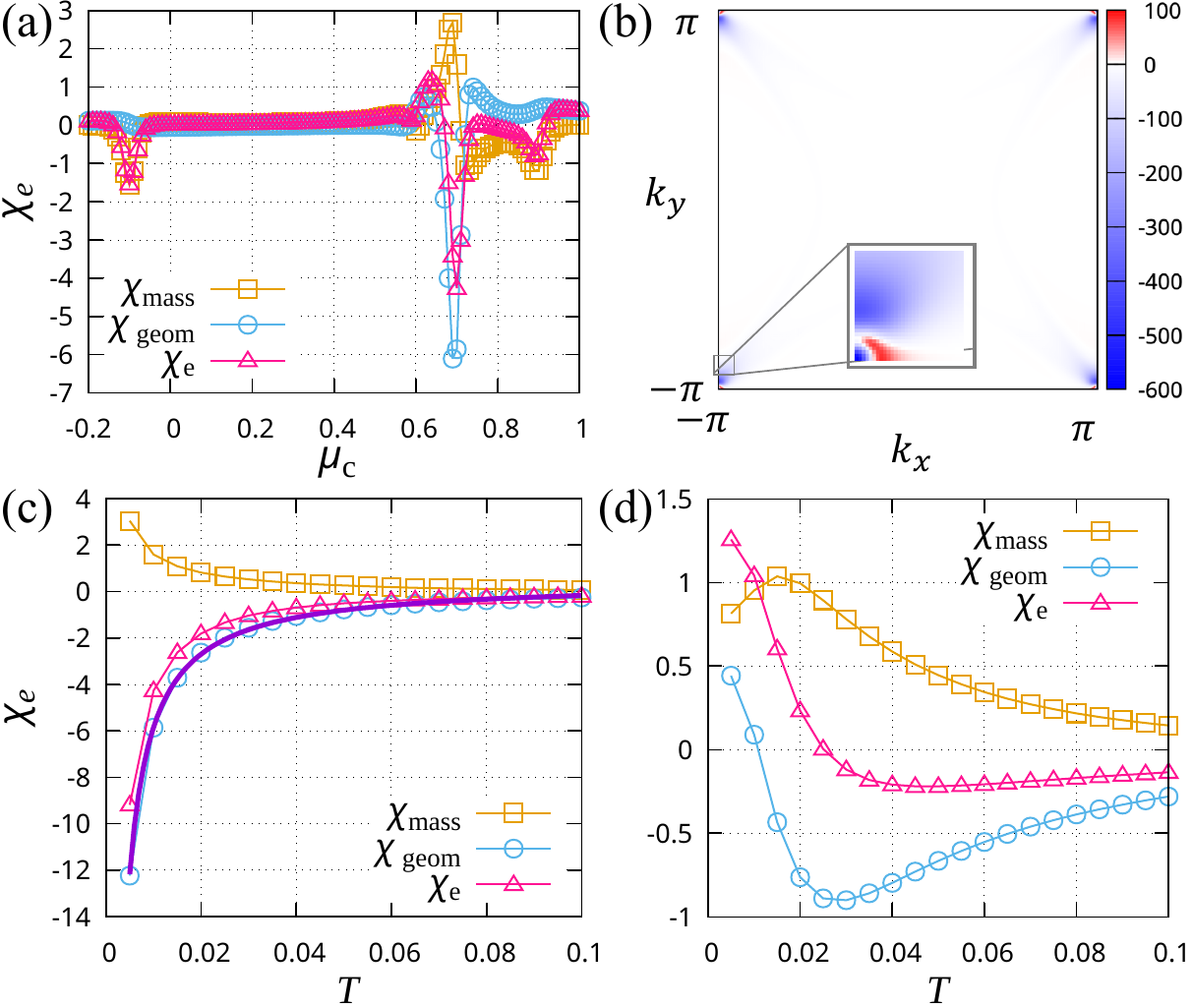}
  \centering
  \caption{GES $\chi_{\rm e}^{0:xx}$ of the dispersive Lieb lattice model. In (a), (c), and (d), the triangles, circles, and squares show $\chi_{\rm e}^{0:xx}$, $\chi_{\rm e:geom}^{0:xx}$, and $\chi_{\rm e:mass}^{0:xx}$, respectively. (a) The $\mu_{\rm c}$ dependence for $T=0.01$. (b) The quantum geometric contribution to the GES from each $\bm k$ point for $(\mu_{\rm c}, T) = (0.7, 0.02)$. The inset shows the contribution near the $M$ point with band degeneracy. (c) and (d) show the temperature dependence for $\mu_{\rm c} = 0.7$ and $\mu_{\rm c} = 0.65$, respectively. The purple line in (c) is a fitting curve $\chi_{\rm e:geom}^{0:xx} \simeq -0.0631779/T + 0.462022$. 
  \label{fig:chie_Lieb_all}}
\end{figure}

In Fig.~\ref{fig:chie_Lieb_all}(a), we show the chemical-potential dependence of GES $\chi_{\rm e}^{0:xx}$. In some regions near the Lifshitz transitions ($\mu_{\rm c} \simeq -0.1$ and $0.9$), the GES shows the dip structure. 
This structure is induced by the effective-mass term.
The effective-mass contribution from each band is proportional to an odd function $f^{(2)}(\epsilon_n(\bm k))$, and therefore, the effective-mass term tends to cancel out between the states below and above the Fermi energy.
However, the cancellation is incomplete for $\mu_{\rm c}$ near the Lifshitz transition point, and thus, the effective-mass term gives a negative GES.
This is an understanding of why ferromagnetic fluctuation appears at finite temperatures when the Fermi surface is small, from the viewpoint of the GES.

In contrast, accompanying the band degeneracy on the Fermi surface, we obtain the maximally negative value of GES $\chi_{\rm e}^{0:xx}$ at $\mu_{\rm c} = 0.7$, which is dominated by the quantum geometric contribution.
As expected from the band degeneracy at the M point, the quantum geometric term of the GES mainly comes from the region near the M point.
This is verified by the $\bm k$-resolved quantum geometric contribution shown in Fig.~\ref{fig:chie_Lieb_all}(b). We find a large negative contribution to the GES from the vicinity of the $M$ point, which in turn induces ferromagnetic fluctuation.

As we have mentioned, the quantum metric gives a negative contribution to the GES, while the positional shift positively contributes. 
Our results imply that the quantum metric overcomes the positional shift when the band-degenerate point lies on the Fermi surface. 
This can be intuitively understood from the formula of the quantum geometric term. 
The quantum metric contributes to the GES with  $f^\prime(\epsilon_n(\bm k))$, which is divergent on the Fermi surface at low temperatures, [$f^\prime(0) \propto 1/T$]. On the other hand, $F_{nm}({\bm k})$ in the positional shift contribution is a regular function. Therefore, the quantum metric becomes significant in the presence of band degeneracy at low energies.
Consistent with the intuitive explanation, the geometric term is negatively enhanced at low temperatures owing to the contribution of quantum metric, as shown in Fig.~\ref{fig:chie_Lieb_all}(c). The geometric term is well fitted by the scaling $\chi_{\rm e:geom}^{0:\mu\nu} = a/T+b$ with constants $a,b$.
Thus, we conclude that the quantum metric on the Fermi surface induces ferromagnetic fluctuation when the non-Kramers band degeneracy lies on the Fermi surface. 

However, when the band-degenerate point is slightly off the Fermi surface and temperature decreases so that $T \ll \vert \mu_{\rm c} -0.7 \vert$, the negative geometric term is suppressed as shown in Fig.~\ref{fig:chie_Lieb_all}(d). 
This is consistent with the fact that the quantum metric contribution is a Fermi-surface term.
As the band-degenerate point moves away from the Fermi surface by much more than $T$, $f^\prime(\epsilon_n(\bm k))$ near the $M$ point decays,
and the quantum metric contribution is suppressed.
At low temperatures, the positional shift contribution overcomes the quantum metric contribution, and the quantum geometric term is positive. Thus, in this case, the ferromagnetic-antiferromagnetic crossover of fluctuation occurs as the temperature decreases.

\begin{figure}[tbp]
  \includegraphics[width=1.0\linewidth]{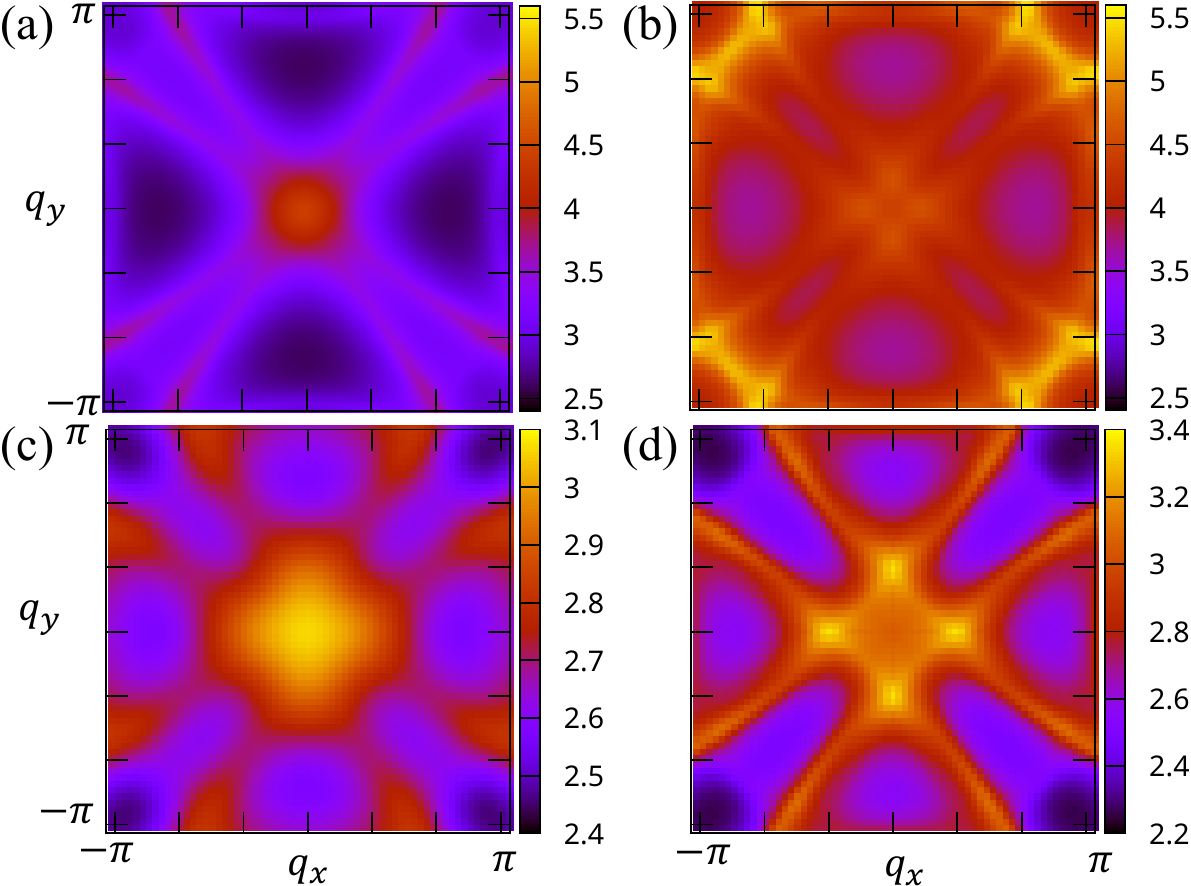}
  \centering
  \caption{The bare spin susceptibility in the dispersive Lieb lattice model. (a) $\chi_{\rm s}^{0}(\bm q)$ and (b) $\chi_{\rm s:\rm band}^{0}(\bm q)$ for $(\mu_{\rm c}, T) = (0.7, 0.01)$ with the same color bar. (c) and (d) show $\chi_{\rm s}^{0}(\bm q)$ for $(\mu_{\rm c}, T) = (0.65, 0.05)$ and $(0.65, 0.01)$, respectively.   \label{fig:chis0_Lieb_all}}
\end{figure}

\textit{Quantum-geometry-induced ferromagnetic fluctuation.---} Then, to justify the above discussion, we show the bare spin susceptibility defined by $\chi_{\rm s}^0(\bm q) = 2\sum_{nm}\int \frac{d\bm k}{(2\pi)^2}F_{nm}(\bm k,\bm q)(1 - D_{nm}(\bm k,\bm q))$, where the quantum distance $D_{nm}(\bm k,\bm q) \equiv 1-\vert\braket{u_n(\bm k+\bm q)\vert u_m(\bm k)}\vert^2$ is closely related to the quantum geometry.
Quantum geometry suppresses $\chi_{\rm s}^0(\bm q)$ at $\bm q\neq0$ via nonzero quantum distance $D_{nn}(\bm k,\bm q)$, which is expanded as $\sim\sum_{\mu\nu} g_{n}^{\mu\nu}(\bm k)q_\mu q_\nu +\cdots$ with the quantum metric. However, $\chi_{\rm s}^0(\bm 0)$ is not suppressed, and ferromagnetic fluctuation is relatively enhanced. The van Vleck susceptibility arising from $D_{nm}(\bm k,\bm q)$ for $n\neq m$ corresponds to the positional-shift contribution to the GES.
For comparison, we also define the bare spin susceptibility without quantum geometry, $\chi_{\rm s:band}^0(\bm q)=2\sum_{n}\int \frac{d\bm k}{(2\pi)^2}F_{nn}(\bm k,\bm q)$,
in which magnetic fluctuation is determined by only the effective-mass term.
By comparing these two quantities, we can elucidate the effects of quantum geometry.

In Figs.~\ref{fig:chis0_Lieb_all}(a) and \ref{fig:chis0_Lieb_all}(b), we show $\chi_{\rm s}^0(\bm q)$ and $\chi_{\rm s:band}^0(\bm q)$ 
in the dispersive Lieb lattice model for $\mu_{\rm c}=0.7$.
As expected by Fig.~\ref{fig:chie_Lieb_all}(a) showing the negative GES, $\chi_{\rm e}^{0:\mu\nu} = \lim_{\bm q=0}\partial_{q_\mu}\partial_{q_\nu}\chi^0(\bm q)$, the bare spin susceptibility shows ferromagnetic fluctuation (Fig.~\ref{fig:chis0_Lieb_all}(a)). However, antiferromagnetic fluctuation is obtained when we neglect the quantum geometry (Fig.~\ref{fig:chis0_Lieb_all}(b)). Thus, we conclude that the quantum geometry induces the ferromagnetic fluctuation.
It is emphasized that the maximum of DOS at $\mu_{\rm c} = 0.7$ is not sufficient for the ferromagnetic fluctuation; the relative enhancement of $\chi_{\rm s}^0(0)$ compared to $\chi_{\rm s}^0(\bm q\neq 0)$ by the quantum distance/quantum geometry is essential. Note that the momentum dependence of spin susceptibility plays an essential role in unconventional superconductivity~\cite{moriya2000spin,yanase2003theory}.
We also show $\chi_{\rm s}^0(\bm q)$ for $(\mu_{\rm c}, T) = (0.65,0.05)$ and $(\mu_{\rm c}, T) = (0.65,0.01)$ in Figs.~\ref{fig:chis0_Lieb_all}(c) and \ref{fig:chis0_Lieb_all}(d), respectively. Consistent with Fig.~\ref{fig:chie_Lieb_all}(d), we confirm the crossover from ferromagnetic to antiferromagnetic fluctuation as the temperature decreases.

\textit{Spin-triplet superconductivity.---}
Finally, we show that quantum-geometry-induced ferromagnetic fluctuation mediates spin-triplet superconductivity. 
To see this, we set the onsite interaction as $U = 0.86$ and solve the linearized gap equation, $\lambda^{\rm t(s)}\Delta_{ll^\prime}(\bm k) = -\frac{1}{N\beta}\sum_{\bm k^\prime\omega_n}\sum_{\{l_i\}}V_{ll_1,l_2l^\prime}^{\rm t(s)}(\bm k-\bm k^\prime)\mathcal{G}_{l_1l_3}(\bm k^\prime,i\omega_n)\Delta_{l_3l_4}(\bm k^\prime)\mathcal{G}_{l_2l_4}(-\bm k^\prime,-i\omega_n)$, using the effective interaction obtained by RPA $V^{\rm t(s)}(\bm q)$, which is $\simeq \frac{-1(3)}{4}U\chi_{\rm s}(\bm q)U$ in single-band systems~\cite{moriya2000spin, yanase2003theory, anderson1973anistropic, miyake1986spin, Scalapino1986} but here extended to multi-band systems~\cite{Supplemental}.
Here, $\mathcal{G}(\bm k,i\omega_n)$ is the Green function with the Matsubara frequency, $i\omega_n$. The instability of spin-triplet (singlet) superconductivity with the form factor $\Delta(\bm k)$ is determined by the maximum eigenvalue $\lambda^{\rm t(s)}$. While the mean-field formalism overestimates the transition temperature, the dynamical effect of effective interaction %non-static effect of $V^{\rm t(s)}(\bm q)$ 
is expected not to alter the superconducting symmetry, as in the cases of  $^3$He~\cite{anderson1973anistropic} and cuprates~\cite{moriya1990antiferromagnetic}.

\begin{figure}[tbp]
  \includegraphics[width=1.0\linewidth]{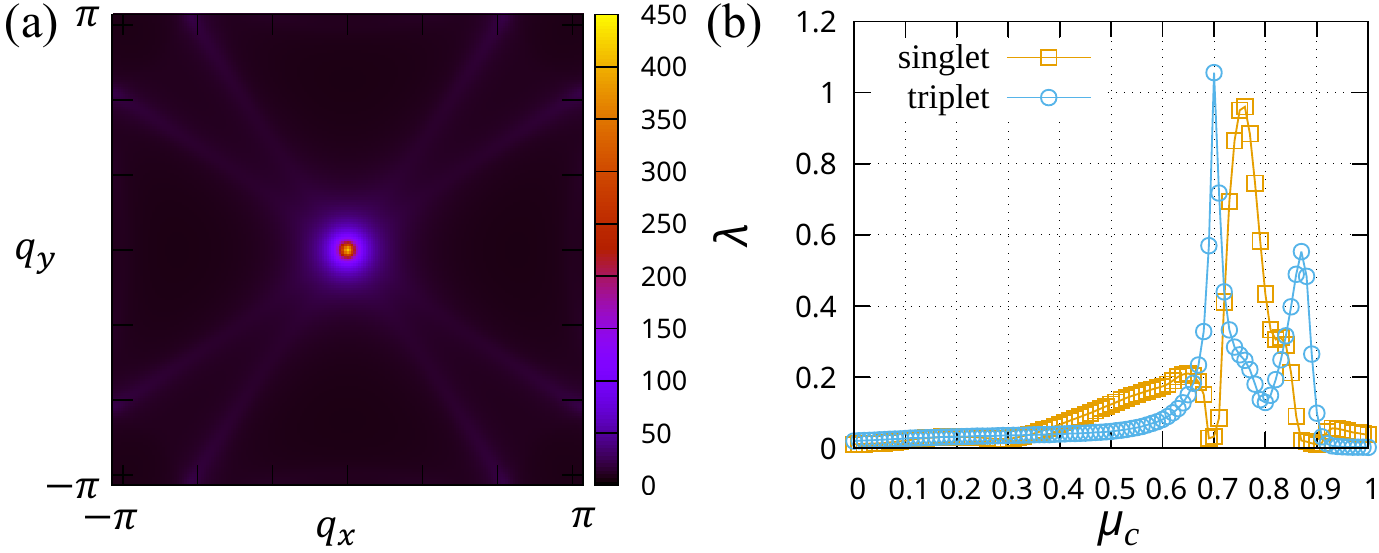}
  \centering
  \caption{(a) The spin susceptibility obtained by RPA for $(\mu_{\rm c}, T) = (0.7, 0.01)$. (b) The eigenvalues of the linearized gap equation at $T = 0.01$. The blue and orange lines show the maximum eigenvalue for spin-triplet and spin-singlet superconductivity, respectively. 
  Eigenvalues for all the irreducible representations are shown in Supplemental Materials~\cite{Supplemental}. \label{fig:sc_Lieb_all}}
\end{figure}

Figure~\ref{fig:sc_Lieb_all}(a) shows the spin susceptibility at $\mu_{\rm c} = 0.7$ obtained by RPA. Ferromagnetic fluctuation is enhanced by the Coulomb interaction, as we see from the comparison to Fig.~\ref{fig:chis0_Lieb_all}(a). Eigenvalues of the linearized gap equation are shown in Fig.~\ref{fig:sc_Lieb_all}(b) for spin-singlet extended-$s$-wave (orange line) and spin-triplet $p$-wave (blue line) superconductivity~\cite{Supplemental}. It is revealed that the spin-triplet superconductivity is stabilized around $\mu_{\rm c} \simeq 0.7$
and $0.9$ corresponding to the negative peak of GES in Fig.~\ref{fig:chie_Lieb_all}(a). 

Especially, we obtain the largest eigenvalue at $\mu_{\rm c} = 0.7$ where quantum geometry induces ferromagnetic fluctuation.
Combined with the large DOS, the strong ferromagnetic fluctuation enhanced by interaction gives a large eigenvalue for spin-triplet superconductivity. Thus, we conclude spin-triplet superconductivity from quantum-geometry-induced ferromagnetic fluctuation.

\textit{Discussion.---}
In this Letter, we show that quantum geometry induces ferromagnetic fluctuation and results in spin-triplet superconductivity. 
The Fubini-Study quantum metric on the Fermi surface is an essential quantity for this mechanism of magnetism and superconductivity.
Using the dispersive Lieb lattice model, we demonstrated that the non-Kramers band degeneracy on the Fermi surface plays the central role in enhancing the quantum-geometry-induced phenomena. 
In the diverse studies on unconventional superconductivity, the quantum geometry of electrons coupled to many-body effects has not been focused on. Stimulated by recent developments in the topology and geometry of quantum materials, we shed light on a route to spin-triplet superconductivity and, thereby, topological superconductivity.

A question of interest is whether our theory can be applied to other systems as well.  
To answer this, we have calculated the GES of Raghu's model~\cite{raghu2008minimal} for iron-based superconductors~\cite{Supplemental}. This model has the non-Kramers band degeneracy at the $\Gamma$ point. 
Also in this model, the quantum geometry induces ferromagnetic fluctuation due to the non-Kramers band degeneracy. 
In addition, we confirmed the quantum-geometry-induced ferromagnetic fluctuation in other models with the flat band and various band touching including the usual Lieb lattice model~\cite{gferrocomment}.
Thus, a wide range of materials with non-Kramers band degeneracy~\cite{Zhang2019,Vergniory2019} are candidates for quantum-geometry-induced ferromagnetism and superconductivity. 
We expect that future material-specific studies will be stimulated by our work. The exploration of two-dimensional materials with high tunability, e.g., by band engineering through heterostructures, gate voltage, strain, and twist angle is also expected.

\begin{acknowledgments}
We are grateful to R. Hakuno, K. Nogaki, Y. Takahashi, T. Nomoto, and R. Arita, for fruitful discussions.
This work was supported by JSPS KAKENHI (Grant Nos. JP18H01178, JP18H05227, JP20H05159, JP21K13880, JP21K18145, JP22H01181, JP22H04476, JP22H04933, JP22J22520).
\end{acknowledgments}

\clearpage

\renewcommand{\bibnumfmt}[1]{[S#1]}
\renewcommand{\citenumfont}[1]{S#1}
\renewcommand{\thesection}{S\arabic{section}}
\renewcommand{\theequation}{S\arabic{equation}}
\setcounter{equation}{0}
\renewcommand{\thefigure}{S\arabic{figure}}
\setcounter{figure}{0}
\renewcommand{\thetable}{S\arabic{table}}
\setcounter{table}{0}
\makeatletter
\c@secnumdepth = 2
\makeatother

\onecolumngrid

\begin{center}
 {\large \textmd{Supplemental Materials:} \\[0.3em]
 {\bfseries Spin-triplet superconductivity from quantum-geometry-induced ferromagnetic fluctuation}}
\end{center}

\begin{center}
\section{Spin, charge, and generalized electric susceptibility}
\end{center}

In this section, we show the detailed calculation of the spin, charge, and generalized electric susceptibility.
While we focus on two-dimensional systems in the main text, the following discussion is written for systems in any dimension, including two and three-dimensional systems.

\begin{center}
\end{center}
\begin{center}
    \subsection{Multi-band Hubbard model with SU(2) symmetry}
\end{center}

We consider the multi-band Hubbard model with SU(2) symmetry, 
\begin{eqnarray}
    \hat{\mathcal{H}} &=& \hat{H}_0 + \hat{H}_{\rm int},\\
    \hat{H}_0 &=& \sum_{\bm k}\sum_{\sigma}\hat{\bm c}_\sigma^\dagger(\bm k)H_0(\bm k)\hat{\bm c}_\sigma(\bm k),\\
    \hat{H}_{\rm int} &=& U\sum_{\bm R}\sum_{l}\hat{ n}_{l\uparrow}(\bm R)\hat{n}_{l\downarrow}(\bm R),
\end{eqnarray}
where $\hat{\bm c}_\sigma^\dagger(\bm k) =( 
\begin{array}{ccccc}
     \hat{c}_{1\sigma}^\dagger(\bm k)&\ldots & \hat{c}_{l\sigma}^\dagger(\bm k)&\ldots & \hat{c}_{f\sigma}^\dagger(\bm k)
\end{array}
)
$ is the creation operator of electrons with the wave vector $\bm k$, spin $\sigma =\uparrow\downarrow$, and the internal degrees of freedom $l$ such as orbitals and sublattices.
The dimension of the internal degrees of freedom
is represented by $f$. $\hat{n}_{l\sigma}(\bm R) = \hat{c}^\dagger_{l\sigma}(\bm R)\hat{c}_{l\sigma}(\bm R)$ is the particle density operator for $l$ and spin $\sigma$ at position $\bm R$. The Fourier transform is defined by $\hat{c}^\dagger_{l\sigma}(\bm R) = \frac{1}{\sqrt{N}}\sum_{\bm k}e^{-i\bm k\cdot(\bm R+\bm r_l)}\hat{c}_{l\sigma}^\dagger(\bm k)$. $H_0(\bm k)$ is the matrix representation of the Fourier transform of hopping integrals with the internal coordinate $\bm r_l$. $U$ is the onsite 
Coulomb interaction, and $N$ is the volume of the system.
The SU(2) symmetry is preserved in this model, since the spin-orbit coupling (SOC) and the magnetic field are absent. We ignore two-body interactions other than the onsite Coulomb interaction, such as an inter-orbital interaction, for simplicity.

\begin{center}
\end{center}

\begin{center}
    \subsection{Spin and charge susceptibility}
\end{center}

The particle density operator for each spin is defined by,
\begin{eqnarray}
    \hat{n}_{\sigma}(\bm q,\tau) &=& \sum_{\bm k}\hat{\bm c}^\dagger_{\sigma}(\bm k,\tau)\hat{\bm c}_{\sigma}(\bm k+\bm q,\tau),
\end{eqnarray}
where $\hat {\bm c}_{\sigma}(\bm k,\tau) = e^{\tau\hat{\mathcal{H}}}\hat {\bm c}_{\sigma}(\bm k)e^{-\tau\hat{\mathcal{H}}}$ is the imaginary-time representation with the imaginary time $\tau$. 
For later calculation, we also define the matrix elements of particle density operators for each spin as,
\begin{eqnarray}
    \hat{n}_{\sigma:ll^\prime}(\bm q,\tau) = \sum_{\bm k}\hat{c}^\dagger_{\sigma l}(\bm k,\tau)\hat{c}_{\sigma l^\prime}(\bm k+\bm q,\tau).
\end{eqnarray}
Using this, the spin susceptibility and the charge susceptibility are defined by,
\begin{eqnarray}
    \chi_{\rm s}(\bm q, i\Omega_n) &=& \chi_{\uparrow\uparrow}(\bm q, i\Omega_n)-\chi_{\uparrow\downarrow}(\bm q, i\Omega_n)-\chi_{\downarrow\uparrow}(\bm q, i\Omega_n)+\chi_{\downarrow\downarrow}(\bm q, i\Omega_n),\notag\\
    &=& 2\chi_{\uparrow\uparrow}(\bm q, i\Omega_n)
    -2\chi_{\uparrow\downarrow}(\bm q, i\Omega_n),\\
    \chi_{\rm c}(\bm q, i\Omega_n) &=& \chi_{\uparrow\uparrow}(\bm q, i\Omega_n)+\chi_{\uparrow\downarrow}(\bm q, i\Omega_n)+\chi_{\downarrow\uparrow}(\bm q, i\Omega_n)+\chi_{\downarrow\downarrow}(\bm q, i\Omega_n),\notag\\
    &=& 2\chi_{\uparrow\uparrow}(\bm q, i\Omega_n)
    +2\chi_{\uparrow\downarrow}(\bm q, i\Omega_n),\label{eq:chi_c_q_iOmega}\\
    \chi_{\sigma\sigma^\prime}(\bm q, i\Omega_n) &=& \dfrac{1}{N}\int_0^{\beta} d\tau e^{i\Omega_n\tau}\braket{T_{\tau}\left[\hat{n}_{\sigma}(\bm q,\tau)\hat{n}_{\sigma^\prime}(-\bm q)\right]},
\end{eqnarray}
with the bosonic Matsubara frequency $i\Omega_n$ and the inverse temperature $\beta$. 
$T_{\tau}$ represents the time-ordering product for $\tau$. Here, we used the relations ensured by the SU(2) symmetry $\chi_{\uparrow\uparrow}(\bm q, i\Omega_n) = \chi_{\downarrow\downarrow}(\bm q, i\Omega_n)$ and $\chi_{\uparrow\downarrow}(\bm q, i\Omega_n) = \chi_{\downarrow\uparrow}(\bm q, i\Omega_n)$.
We also define the matrix representation $\bar{\chi}_{\sigma\sigma^\prime}(\bm q, i\Omega_n)$ of $\chi_{\sigma\sigma^\prime}(\bm q, i\Omega_n)$ using the matrix element written by, 
\begin{eqnarray}
    \left[\bar{\chi}_{\sigma\sigma^\prime}(\bm q, i\Omega_n)\right]_{l_1l_1^\prime,l_2l_2^\prime} = \dfrac{1}{N}\int_0^{\beta} d\tau e^{i\Omega_n\tau}\braket{T_{\tau}\left[\hat{n}_{\sigma:\l_1^\prime\l_1}(\bm q,\tau)\hat{n}_{\sigma':l_2l_2^\prime}(-\bm q)\right]}.
\end{eqnarray}

\begin{center}
\end{center}
\begin{center}    \subsubsection{Noninteracting system}
\end{center}

The spin (charge) susceptibility of a noninteracting system, namely the bare spin (charge) susceptibility, can be written as,
\begin{eqnarray}
    \chi_{\rm s}^0(\bm q, i\Omega_n) &=& \chi_{\rm c}^0(\bm q, i\Omega_n) = 2\chi^0(\bm q, i\Omega_n),\\
    \chi^0(\bm q, i\Omega_n) &=& \chi^0_{\sigma\sigma}(\bm q, i\Omega_n)
    =\dfrac{1}{N}\int_0^{\beta} d\tau e^{i\Omega_n\tau}\braket{T_{\tau}\left[\hat{n}_{\sigma}(\bm q,\tau)\hat{n}_{\sigma}(-\bm q)\right]},\notag\\
    &=& -\dfrac{1}{N\beta}\sum_{\bm k\omega_n}{\rm Tr}\left[\mathcal{G}(\bm k+\bm q,i\omega_n+i\Omega_n)\mathcal{G}(\bm k,i\omega_n)\right].\label{eq:chi0_q_iOmegan}
\end{eqnarray}
We used the property of noninteracting systems, $\chi^0_{\uparrow\downarrow}(\bm q, i\Omega_n) = \chi^0_{\downarrow\uparrow}(\bm q, i\Omega_n) = 0$, which is satisfied by the SU(2) symmetry of Hamiltonian. Here, we define the Green function of noninteracting systems $\mathcal{G}(\bm k,i\omega_n) = [i\omega_n-H_0(\bm k)]^{-1}$ with fermionic Matsubara frequency $\omega_n$. Tr represents the trace for all degrees of freedom except for the spin. 
The matrix elements are obtained as,
\begin{eqnarray}
    \dfrac{1}{2}[\bar{\chi}_{\rm s(c)}^0(\bm q, i\Omega_n)]_{l_1l_1^\prime,l_2l_2^\prime}&=&
    [\bar{\chi}^0(\bm q, i\Omega_n)]_{l_1l_1^\prime,l_2l_2^\prime} \notag\\
    &=& -\dfrac{1}{N\beta}\sum_{\bm k\omega_n}\left[\mathcal{G}_{l_1l_2}(\bm k+\bm q,i\omega_n+i\Omega_n)\mathcal{G}_{l_2^\prime l_1^\prime}(\bm k,i\omega_n)\right].
\end{eqnarray}
After taking the sum of Matsubara frequency, we get,
\begin{eqnarray}
    \chi^0(\bm q, i\Omega_n)= \sum_{nm}\int \dfrac{d\bm k}{(2\pi)^d}\dfrac{f(\epsilon_n(\bm k+\bm q))-f(\epsilon_m(\bm k))}{\epsilon_m(\bm k)-\epsilon_n(\bm k+\bm q)+i\Omega_n}\vert\braket{u_n(\bm k+\bm q)\vert u_m(\bm k)}\vert^2,
\end{eqnarray}
with the dimension of the system $d$.
We can calculate the band dispersion $\epsilon_n(\bm k)$ and Bloch wave function $\ket{u_n(\bm k)}$ by the eigenvalue equation, $H_0(\bm k)\ket{u_n(\bm k)} = \epsilon_n(\bm k) \ket{u_n(\bm k)}$.

\begin{center}
\end{center}
\begin{center}
    \subsubsection{Random phase approximation}
\end{center}

We define the irreducible vertex,
\begin{eqnarray}
    \left[\Gamma^0\right]_{l_1l_1^\prime,l_2l_2\prime} = U\delta_{l_1l_1^\prime}\delta_{l_1^\prime l_2}\delta_{l_2l_2^\prime}.
\end{eqnarray}
In the random phase approximation (RPA), the spin (charge) susceptibility and its matrix representation are given by, 
\begin{eqnarray}
    &&\chi_{\rm s(c)}(\bm q,i\Omega_n) = \sum_{ll^\prime}
    \left[\bar{\chi}_{\rm s(c)}(\bm q,i\Omega_n)\right]_{ll,l^\prime l^\prime}
    \notag\\
    &&\bar{\chi}_{\rm s(c)}(\bm q,i\Omega_n) =  \left[1\mp\Gamma_0\bar{\chi}^0(\bm q,i\Omega_n)\right]^{-1}\bar{\chi}_{\rm s(c)}^0(\bm q,i\Omega_n).
\end{eqnarray}

\begin{center}
\end{center}

\begin{center}
\subsection{Generalized electric susceptibility}
\end{center}

In this subsection, we derive the generalized electric susceptibility using the Kubo formula and local thermodynamics.

\begin{center}
\end{center}

\begin{center}
\subsubsection{Electric and charge susceptibility via Kubo Formula}
\end{center}

First, we show an alternative way to introduce charge susceptibility by using linear response theory.
Based on the Kubo formula, the charge susceptibility of real-space and real-time representation $\chi_{\rm c}(\bm r, t)$ with a position $\bm r$ and real time $t$ is defined by, 
\begin{eqnarray}
\braket{\hat{n}(\bm r, t)} = \rho_0-\int d\bm r^\prime\int_{-\infty}^{t} dt^\prime \chi_{\rm c}(\bm r-\bm r^\prime,t-t^\prime)\phi(\bm r^\prime,t^\prime).
\end{eqnarray}
Here, $\braket{\hat{n}(\bm r, t)}$ is the expectation value of the particle density operator, $\hat{n}(\bm r, t) = \sum_{l,\sigma}\hat{c}_{l, \sigma}^\dagger(\bm r, t)\hat{c}_{l, \sigma}(\bm r, t)$, namely the charge density, $\phi(\bm r,t)$ is an external scalar potential, and $\rho_0$ is the charge density in the absence of the external field.
Also, $\hat{c}_{l, \sigma}^\dagger(\bm r, t) = e^{i\hat{\mathcal{H}}t/\hbar}\hat{c}_{l,\sigma}^\dagger(\bm r)e^{-i\hat{\mathcal{H}}t/\hbar}$ is the Heisenberg representation of creation operator with the Dirac constant $\hbar=1$ for the natural unit. 
After the Fourier transform with respect to real time, we get the frequency representation of the charge susceptibility,
\begin{eqnarray}
  \braket{\hat{n}(\bm r, t)} &=& \rho_0-\int d\bm r^\prime\int_{-\infty}^{\infty} \dfrac{d\omega}{2\pi}e^{-i\omega t+\delta t}\chi_{\rm c}(\bm r-\bm r^\prime,\omega)\phi(\bm r^\prime,\omega),\\
  \chi_{\rm c}(\bm r,\omega) &=& \int_{-\infty}^{t} dt^\prime e^{i\omega(t-t^\prime)-\delta(t-t^\prime)}\chi_{\rm c}(\bm r,t-t^\prime),\\
  \phi(\bm r,t) &=& \int_{-\infty}^{\infty} \dfrac{d\omega}{2\pi} e^{-i\omega t+\delta t}\phi(\bm r,\omega).
\end{eqnarray}
Here, the infinitesimal $\delta$ ensures that the external field vanishes at $t\rightarrow-\infty$.
We can also define the frequency representation of the charge density as,
\begin{eqnarray}
    \braket{\hat{n}(\bm r)}(\omega) &=& \delta(\omega)\rho_0-\int d\bm r^\prime\chi_{\rm c}(\bm r-\bm r^\prime,\omega)\phi(\bm r^\prime,\omega),~\label{eq:n_r_omega}\\
    \braket{\hat{n}(\bm r,t)} &=& \int_{-\infty}^\infty\dfrac{d\omega}{2\pi}e^{-i\omega t+\delta t}\braket{\hat{n}(\bm r)}(\omega).
\end{eqnarray}

Since the correlation function should decay away from the external field at $\bm r$, the integrand of Eq.~\eqref{eq:n_r_omega} contributes only when $\bm r-\bm r^\prime$ is sufficiently small. In contrast, we assume that the scalar potential spatially modulates on a length scale larger than that of the correlation function. Therefore, we can expand the scalar potential by $\bm r-\bm r^\prime$ as,
\begin{eqnarray}
    \braket{\hat{n}(\bm r)}(\omega) -\delta(\omega)\rho_0 &=& -\int d\bm r^\prime\chi_{\rm c}(\bm r-\bm r^\prime,\omega)\phi(\bm r-(\bm r-  \bm r^\prime),\omega),\notag\\
    &=& -\sum_{n=0}^{\infty}\int d\bm r^\prime\chi_{\rm c}(\bm r-\bm r^\prime,\omega)\dfrac{\left[(-r_{\mu}+r^\prime_{\mu})\partial_{r_{\mu}}\right]^n}{n!}\phi(\bm r,\omega),\notag\\
    &=& - \dfrac{1}{N}\sum_{\bm q}\sum_{n=0}^{\infty}\int d\bm r^\prime e^{i\bm q\cdot(\bm r-\bm r^\prime)}\chi_{\rm c}(\bm q,\omega)\dfrac{\left[(-r_{\mu}+r^\prime_{\mu})\partial_{r_{\mu}}\right]^n}{n!}\phi(\bm r,\omega),\notag\\
    &=& - \sum_{n=0}^{\infty}\int\dfrac{d\bm q}{(2\pi)^d}\int d\bm r^\prime e^{i\bm q\cdot(\bm r-\bm r^\prime)}\left[-i\partial_{q_{\mu}}\partial_{r_{\mu}}\right]^n\chi_{\rm c}(\bm q,\omega)\dfrac{\phi(\bm r,\omega)}{n!},\notag\\
    &=& -\sum_{n=0}^{\infty}\lim_{\bm q\rightarrow 0}\left[-i\partial_{q_{\mu}}\partial_{r_{\mu}}\right]^n\chi_{\rm c}(\bm q,\omega)\dfrac{\phi(\bm r,\omega)}{n!},\notag\\\label{eq:n-rho}
\end{eqnarray}
where and hereafter, we take the sum of repeated indices, such as $\mu = x,y,z$. For example, Eq.~\eqref{eq:n-rho} up to $n = 2$ is explicitly written as, 
\begin{eqnarray}
    \braket{\hat{n}(\bm r)}(\omega) -\delta(\omega)\rho_0 
    &=& -\chi_{\rm c}(0,\omega)\phi(\bm{r},\omega)+i\lim_{\bm q\rightarrow0}\sum_{\mu_1}\partial_{q_{\mu}}\chi_{\rm c}(\bm q,\omega)\partial_{r_{\mu}}\phi(\bm r,\omega)\notag\\
    &+&\dfrac{1}{2}\lim_{\bm q\rightarrow0}\sum_{\mu\nu}\partial_{q_{\mu}}\partial_{q_{\nu}}\chi_{\rm c}(\bm q,\omega)\partial_{r_{\mu}}\partial_{r_{\nu}}\phi(\bm r,\omega).
\end{eqnarray}
Through the analytic continuation $\omega +i\delta\rightarrow i\Omega_n$, $\chi_{\rm c}(\bm q,\omega)$ corresponds to Eq.~\eqref{eq:chi_c_q_iOmega}.

Here, we focus on $\chi_{\rm c}(\bm q,\omega = 0)$ for which the system does not depend on time. In other words, we focus on the particle density in equilibrium. When the system is metal and/or at a finite temperature, an equilibrium charge with an external electric field cannot be defined, since the electric current follows in metals or at finite temperatures. Therefore, we consider an insulator at zero temperature. Considering the Lehmann representation
\footnote{The Lehmann representation of charge susceptibility is written by
\begin{eqnarray}
    \chi_{\rm c}(\bm q,\omega) = -\dfrac{1}{N\mathcal{Z}}\sum_{ab}\left(e^{-\beta E_a}-e^{-\beta E_b}\right)\dfrac{\bra{a}\hat{n}(\bm q)\ket{b}\bra{b}\hat{n}(-\bm q)\ket{a}}{\hbar\omega+E_a-E_b+i\delta}.
\end{eqnarray}
Here, $\mathcal{Z}$ is the partition function. $\ket{a}$ and $E_a$ are the eigenstate and the eigenvalue of the many-body Hamiltonian $\hat{\mathcal{H}}$.
}, we see that the charge susceptibility satisfies the relationship $\chi_{\rm c}(\bm q,\omega) = \chi_{\rm c}^{*}(-\bm q,-\omega)$ which means $\chi_{\rm c}(\bm q) = \chi_{\rm c}(-\bm q) (=\chi_{\rm c}(\bm q,0)) $. Therefore, odd-order derivatives of $\chi_{\rm c}(\bm q,0)$ with respect to $\bm q$ vanish. As a result, up to the second order of $\partial_{r_\mu}$, the equilibrium charge density is obtained as, 
\begin{eqnarray}
    \braket{\hat{n}(\bm r)} -\rho_0 &=& -\chi_{\rm c}(0)\phi(\bm r)+\dfrac{1}{2}\lim_{\bm q\rightarrow 0}\partial_{q_\mu}\partial_{q_\nu}\chi_{\rm c}(\bm q)\partial_{r_\mu}\partial_{r_\nu}\phi(\bm r),\notag\\
    &=& -\chi_{\rm c}(0)\phi(\bm r)-\partial_{r_\mu}\left(\dfrac{1}{2}\lim_{\bm q\rightarrow 0}\partial_{q_\mu}\partial_{q_\nu}\chi_{\rm c}(\bm q)E_{\nu}(\bm r)\right),\label{eq:electric}
\end{eqnarray}
where $E_\nu(\bm r) = \partial_{r_\nu}\phi(\bm r)$ is the external electric field.
In the insulator at zero temperature, the first term vanishes and this directly means that $\dfrac{1}{2}\lim_{\bm q\rightarrow 0}\partial_{q_\mu}\partial_{q_\nu}\chi_{\rm c}(\bm q)$ is the electric susceptibility which gives the correction to the charge density, $\delta\braket{\hat{n}(\bm r)} = \braket{\hat{n}(\bm r)} -\rho_0$.

\begin{center}
\end{center}

\begin{center}
\subsubsection{Generalized electric susceptibility via local thermodynamics}
\end{center}

Next, to clarify the physical meaning of the quantity $\dfrac{1}{2}\lim_{\bm q\rightarrow 0}\partial_{q_\mu}\partial_{q_\nu}\chi_{\rm c}(\bm q)$ in metals and/or at a finite temperature, we use the local thermodynamics~\cite{Sshitade2018theory,Sshitade2019theory,Sgao2018microscopic,Sdaido2020thermodynami,Skitamura2021thermodynamic}.
The following discussion is based on Ref.~\onlinecite{Sdaido2020thermodynami}.
We consider the scalar potential arising from an inhomogeneous distribution of disorders, structural asymmetry, contact with a substrate, and so on, rather than an applied electric field. The electric field is assumed to be in a small region near $\bm r$ compared to the volume $N$. In this setup, while the local particle number depends on $\bm r$ due to the spatial variation of $\phi(\bm{r})$, the total particle number is assumed to be constant. Thus, the system is static and the charge current does not flow. The setup is discussed in more detail in Ref.~\onlinecite{Sdaido2020thermodynami}. 

The length scale of $\phi(\bm r)$ is sufficiently longer than the decay length of the Green function. Therefore, $\phi(\bm r)$ varies slowly in space and the system around $\bm r$ is well approximated by the uniform Hamiltonian in which chemical potential $\mu_{\rm c}$ is replaced by $\mu_{\rm c} - \phi(\bm r)$.
In local thermodynamics, starting from the above assumption, the Hamiltonian is expanded by $\bm r^\prime -\bm r$ through $\phi(\bm r^\prime)$ around $\bm r^\prime = \bm r$. 
As a result, free energy depends on $\bm r$ through $\phi(\bm r)$ and is expanded as ,
\begin{align}
    F(\bm r) =& F_0(\mu_{\rm c}-\phi(\bm r)) + Q_{\mu\nu}(\mu_{\rm c}-\phi(\bm r))\partial_{r_\mu}\partial_{r_\nu}\phi(\bm r) \notag\\
    -& \dfrac{1}{2}\partial_{\mu_{\rm c}} Q_{\mu\nu}(\mu_{\rm c}-\phi(\bm r))\partial_{r_\mu}\phi(\bm r)\partial_{r_\nu}\phi(\bm r) + \mathcal{O}(ql)^3,\label{eq:f_phi}
\end{align}
where $F_0(\mu_{\rm c}-\phi(\bm r))$ is the free energy of uniform Hamiltonian with chemical potential $\mu_{\rm c}-\phi(\bm r)$.
In Eq.~\eqref{eq:f_phi}, $Q_{\mu\nu}$ is the thermodynamic electric quadrupole moment defined in Ref.~\onlinecite{Sdaido2020thermodynami}.

The charge density is defined by,
\begin{eqnarray}
    \braket{\hat{n}(\bm r)} &=& -\partial_{\mu_{\rm c}} F(\bm r)\notag\\
    &=&\rho_0(\mu_{\rm c}-\phi(\bm r)) + \partial_{r_\mu}\partial_{r_\nu}Q_{\mu\nu}(\mu_{\rm c}-\phi(\bm r)) -\dfrac{1}{2}\partial_{\mu_{\rm c}}^2Q_{\mu\nu}(\mu_{\rm c}-\phi(\bm r))\partial_{r_\mu}\phi(\bm r)\partial_{r_\nu}\phi(\bm r),\notag\\
\end{eqnarray}
where $\rho_0(\mu_{\rm c}-\phi(\bm r)) = -\partial_{\mu_{\rm c}}F_0(\mu_{\rm c}-\phi(\bm r))$ is the charge density with the chemical potential $\mu_{\rm c}-\phi(\bm r)$.
Therefore, in the linear response theory, this can be written as,
\begin{eqnarray}
    \braket{\hat{n}(\bm r)} -\rho_0 = -\chi_{\rm c}(0)\phi(\bm r)
    -\partial_{r_\mu}\left(\partial_{\mu_{\rm c}}Q_{\mu\nu}E_\nu(\bm r)\right)\label{eq:n_eqm}.
\end{eqnarray}
The first term is the charge susceptibility since it is also defined by $\chi_{\rm c}(0) = -\lim_{\phi(\bm r)\rightarrow0}\dfrac{\delta\rho(\mu-\phi(\bm r))}{\delta\phi(\bm r)}$.
Comparing Eqs.~\eqref{eq:electric} and \eqref{eq:n_eqm},  we get the relationship,
\begin{eqnarray}
    \dfrac{1}{2}\lim_{\bm q\rightarrow 0}\partial_{q_\mu}\partial_{q_\nu}\chi_{\rm c}(\bm q) = \partial_{\mu_{\rm c}}Q_{\mu\nu}.
\end{eqnarray}
Therefore, $\dfrac{1}{2}\lim_{\bm q\rightarrow 0}\partial_{q_\mu}\partial_{q_\nu}\chi_{\rm c}(\bm q)$ is a thermodynamic quantity even in metals and at a finite temperature, and we call it generalized electric susceptibility as a naive generalization of the electric susceptibility to metals.

\begin{center}
\end{center}

\begin{center}
\subsubsection{Derivation of generalized electric susceptibility in noninteracting systems}
\end{center}

We derive the formula of the generalized electric susceptibility in noninteracting systems. Starting from Eq.~\eqref{eq:chi0_q_iOmegan} at $i\Omega_n = 0$, the generalized electric susceptibility is written by,
\begin{eqnarray}
    \lim_{\bm q\rightarrow0}\partial_{q_\mu}\partial_{q_\nu}\chi^0(\bm q) &=& \dfrac{1}{N\beta}\sum_{\bm k\omega_n}{\rm Tr}\left[\partial_{k_\mu}\mathcal{G}(\bm k,i\omega_n)\partial_{k_\nu}\mathcal{G}(\bm k,i\omega_n)\right],\notag\\
    &=& \dfrac{1}{N\beta}\sum_{\bm k\omega_n}\sum_{nm}\dfrac{\bra{u_n(\bm k)}\partial_{k_\mu}H_0(\bm k)\ket{u_m(\bm k)}}{\left(i\omega_n-\epsilon_n(\bm k)\right)^2}\dfrac{\bra{u_m(\bm k)}\partial_{k_\nu}H_0(\bm k)\ket{u_n(\bm k)}}{\left(i\omega_n-\epsilon_m(\bm k)\right)^2},\notag\\
    &=& \dfrac{1}{N}\sum_{\bm k}\sum_{n}\dfrac{f^{(3)}(\epsilon_n(\bm k))}{6}\bra{u_n(\bm k)}\partial_{k_\nu}H_0(\bm k)\ket{u_n(\bm k)}\bra{u_n(\bm k)}\partial_{k_\nu}H_0(\bm k)\ket{u_n(\bm k)}\notag\\
    &+&\dfrac{1}{N}\sum_{\bm k}\sum_{n\neq m}\left(f^\prime(\epsilon_n(\bm k))+f^\prime(\epsilon_m(\bm k))+2\dfrac{f(\epsilon_m(\bm k))-f(\epsilon_n(\bm k))}{\epsilon_n(\bm k)-\epsilon_m(\bm k)}\right)\notag\\
    &\times&\dfrac{\bra{u_n(\bm k)}\partial_{k_\mu}H_0(\bm k)\ket{u_m(\bm k)}}{\left(\epsilon_n(\bm k)-\epsilon_m(\bm k)\right)}\dfrac{\bra{u_m(\bm k)}\partial_{k_\nu}H_0(\bm k)\ket{u_n(\bm k)}}{\left(\epsilon_n(\bm k)-\epsilon_m(\bm k)\right)}.
\end{eqnarray}
The first term is the effective-mass term $\chi_{\rm e:mass}^{0:\mu\nu}$ while the second term is the quantum geometric term $\chi_{\rm e:geom}^{0:\mu\nu}$. Note that, when two bands are degenerate, contribution to the quantum geometric term from the two degenerated bands $\tilde{n},\tilde{m}$ at ${\bm k}$ is
\begin{eqnarray}
    \chi_{{\rm e:geom}:\tilde{n}\tilde{m}}^{0:\mu\nu} = \dfrac{f^{(3)}(\epsilon_{\tilde{n}}(\bm k))}{6}\bra{u_{\tilde{n}}(\bm k)}\partial_{k_\nu}H_0(\bm k)\ket{u_{\tilde{m}}(\bm k)}\bra{u_{\tilde{m}}(\bm k)}\partial_{k_\nu}H_0(\bm k)\ket{u_{\tilde{n}}(\bm k)}+c.c.
\end{eqnarray}
Thus, the geometric term does not diverge even in the presence of band touching.

Then, by using the Hellmann-Feynman theorem,
\begin{eqnarray}
    \bra{u_n(\bm k)}\partial_{k_\mu}H_0(\bm k)\ket{u_m(\bm k)} = \delta_{nm}\partial_{k_\mu}\epsilon_n(\bm k) + \left(\epsilon_n(\bm k)-\epsilon_m(\bm k)\right)\braket{\partial_{k_\mu}u_n(\bm k)\vert u_m(\bm k)},
\end{eqnarray}
the effective-mass and quantum geometric terms are rewritten as 
\begin{eqnarray}
    \chi_{\rm e:mass}^{0:\mu\nu} &=& \dfrac{1}{N}\sum_{\bm k}\sum_{n}\dfrac{f^{(3)}(\epsilon_n(\bm k))}{6}\partial_{k_\mu}\epsilon_n(\bm k)\partial_{k_\nu}\epsilon_n(\bm k),\notag\\
    &=&-\sum_{n}\int \dfrac{d\bm k}{(2\pi)^d}\dfrac{f^{(2)}(\epsilon_n(\bm k))}{6}\partial_{k_\nu}\partial_{k_\mu}\epsilon_n(\bm k),\\
    \chi_{\rm e:geom}^{0:\mu\nu} &=& \dfrac{1}{N}\sum_{\bm k}\sum_{n\neq m}\left(f^\prime(\epsilon_n(\bm k))+f^\prime(\epsilon_m(\bm k))+2\dfrac{f(\epsilon_m(\bm k))-f(\epsilon_n(\bm k))}{\epsilon_n(\bm k)-\epsilon_m(\bm k)}\right)\notag\\
    &&\times\braket{\partial_{k_\mu}u_n(\bm k)\vert u_m(\bm k)}\braket{u_m(\bm k)\vert \partial_{k_\nu}u_n(\bm k)},\notag\\
    &=&\sum_{n}\int \dfrac{d\bm k}{(2\pi)^d}\left(f^\prime(\epsilon_n(\bm k))\sum_{m(\neq n)}\braket{\partial_{k_\mu}u_n(\bm k)\vert u_m(\bm k)}\braket{u_m(\bm k)\vert \partial_{k_\nu}u_n(\bm k)}+c.c.\right.\notag\\
    &&+2\left.f(\epsilon_n(\bm k))\sum_{m(\neq n)}\dfrac{\braket{\partial_{k_\mu}u_n(\bm k)\vert u_m(\bm k)}\braket{u_m(\bm k)\vert \partial_{k_\nu}u_n(\bm k)}+c.c.}{\epsilon_m(\bm k)-\epsilon_n(\bm k)}\right).
\end{eqnarray}
Here, $\partial_{k_\nu}\partial_{k_\mu}\epsilon_n(\bm k)$ is the effective mass, while $\sum_{m(\neq n)}\braket{\partial_{k_\mu}u_n(\bm k)\vert u_m(\bm k)}\braket{u_m(\bm k)\vert \partial_{k_\nu}u_n(\bm k)}+c.c$ and $\sum_{m(\neq n)}(\braket{\partial_{k_\mu}u_n(\bm k)\vert u_m(\bm k)}\braket{u_m(\bm k)\vert \partial_{k_\nu}u_n(\bm k)}+c.c.)/(\epsilon_m(\bm k)-\epsilon_n(\bm k))$ are the quantum metric and the positional shift, respectively. Note that this formula is equivalent to the formula derived by $\partial_{\mu_{\rm c}}Q_{\mu\nu}$ of noninteracting systems.

As for the contribution from the quantum metric,
$f^\prime(\epsilon_n(\bm k))$ is negative and the quantum metric has a positive value. Therefore, the quantum metric always gives a negative contribution to the generalized electric susceptibility.
In contrast, the contribution from the positional shift can be rewritten as,
\begin{eqnarray}
    \chi_{\rm shift}^{0:\mu\nu} &=& \sum_{\bm k}\sum_{n\neq m}\int \dfrac{d\bm k}{(2\pi)^d}\dfrac{f(\epsilon_m(\bm k))-f(\epsilon_n(\bm k))}{\epsilon_n(\bm k)-\epsilon_m(\bm k)}\left(\braket{\partial_{k_\mu}u_n(\bm k)\vert u_m(\bm k)}\braket{u_m(\bm k)\vert \partial_{k_\nu}u_n(\bm k)}+c.c\right).\notag\\
\end{eqnarray}
Since the band resolved quantum metric, $\braket{\partial_{k_\mu}u_n(\bm k)\vert u_m(\bm k)}\braket{u_m(\bm k)\vert \partial_{k_\nu}u_n(\bm k)}+c.c$, and the Lindhard function, $(f(\epsilon_m(\bm k))-f(\epsilon_n(\bm k)))/(\epsilon_n(\bm k)-\epsilon_m(\bm k))$, are positive, this contribution is always positive.

\begin{center}
\end{center}

\begin{center}
\subsubsection{Generalized electric susceptibility of isotropic continuum model in two dimension}
\end{center}
We consider the two-dimensional isotropic continuum model whose Hamiltonian is given by
\begin{eqnarray}
    H(\bm k) = \epsilon(\bm k)-\mu_{\rm c} = \sum_{\mu = x,y}\dfrac{k_\mu^2}{2m}-\mu_{\rm c}.
\end{eqnarray}
In this model, the generalized electric susceptibility can be written as,
\begin{eqnarray}
    \chi_{e}^{0:\mu\mu}
    &=& -\int\dfrac{d\bm k}{(2\pi)^2}\dfrac{f^{(2)}\left(\epsilon(\bm k)-\mu_{\rm c}\right)}{6}\dfrac{1}{m},\notag\\
    &=& -\int_{-\mu_{\rm c}}^\infty d\epsilon\dfrac{f^{(2)}\left(\epsilon\right)}{6}\dfrac{1}{m}\dfrac{m}{2\pi},\notag\\
    &=& -\left[\dfrac{f^\prime\left(\epsilon\right)}{12\pi}\right]_{-\mu_{\rm c}}^\infty=\dfrac{f^\prime\left(\mu_{\rm c}\right)}{12\pi},
\end{eqnarray}
since the density of states is constant, $m/2\pi$. 
Note that the geometric term is absent in the single-band model.
Therefore, the generalized electric susceptibility $\chi_{e}^{0:\mu\mu}$ vanishes at low temperatures which satisfy $\mu_{\rm c} \gg T$; ferromagnetic fluctuation is prohibited in two-dimensional isotropic continuum models. 

\begin{center}
\end{center}

\begin{center}
\section{Dispersive Lieb lattice model}
\end{center}

\begin{center}
\end{center}

\begin{center}
\subsection{Hamiltonian}
\end{center}

\begin{figure}[htbp]
    \centering
    \includegraphics[width=1.0\linewidth]{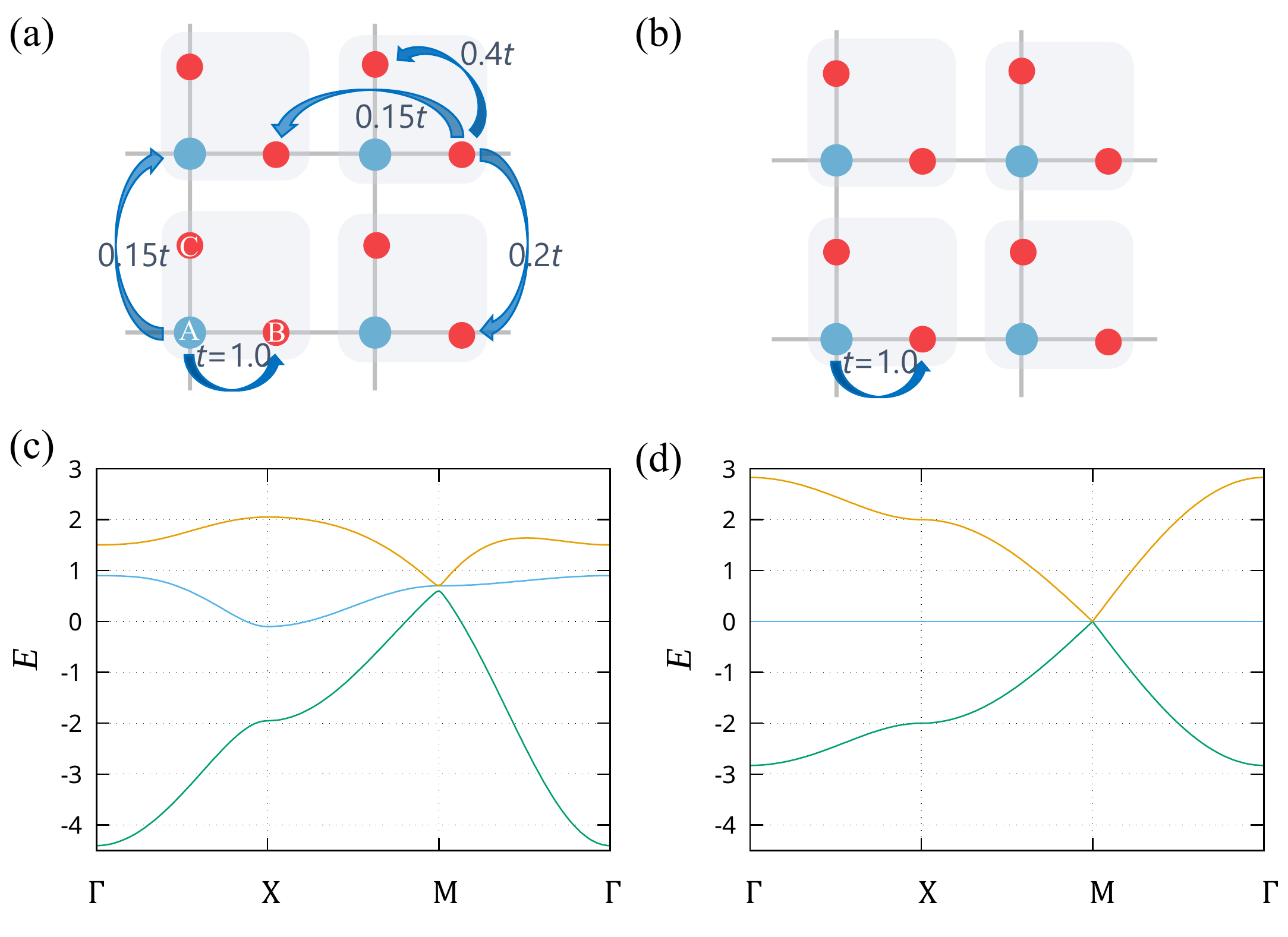}
    \caption{(a) The dispersive Lieb lattice and (b) the usual Lieb lattice. The unit cell (gray box) and three sublattices A, B, and C are shown. Hopping integrals (blue arrows) are illustrated with a unit of the nearest-neighbor hopping. (c) and (d) show the band dispersion of the dispersive and usual Lieb lattice models, respectively.}
    \label{fig:hk_lieb_all2}
\end{figure}

We introduce the Hamiltonian of the dispersive Lieb lattice model. For comparison, we also show the usual Lieb lattice model. In Figs.~\ref{fig:hk_lieb_all2}(a) and \ref{fig:hk_lieb_all2}(b), the hopping integrals of the dispersive and usual Lieb lattice models are schematically shown. In contrast to the usual Lieb lattice model, the dispersive Lieb lattice model includes second- and third-nearest-neighbor hopping integrals. Thus, the noninteracting Hamiltonian for the dispersive Lieb lattice model and the usual Lieb lattice model are written as
\begin{eqnarray}
    H_{0:{\rm d}}(\bm k) &=& \left(
    \begin{array}{ccc}
        \epsilon_{\rm A}(\bm k)-\mu_{\rm c} & \epsilon_{\rm x}(\bm k) & \epsilon_{\rm y}(\bm k)\\
        \epsilon_{\rm x}(\bm k) & \epsilon_{\rm B}(\bm k)-\mu_{\rm c} & \epsilon_{\rm xy}(\bm k)\\
        \epsilon_{\rm y}(\bm k) & \epsilon_{\rm xy}(\bm k) & \epsilon_{\rm C}(\bm k)-\mu_{\rm c}
    \end{array}
    \right),~\label{eq:dispersive_lieb}\\
    H_{0:{\rm l}}(\bm k) &=& \left(
    \begin{array}{ccc}
        -\mu_{\rm c} & \epsilon_{\rm x}(\bm k) & \epsilon_{\rm y}(\bm k)\\
        \epsilon_{\rm x}(\bm k) &-\mu_{\rm c} & 0\\
        \epsilon_{\rm y}(\bm k) & 0 & -\mu_{\rm c}
    \end{array}
    \right),
\end{eqnarray}
respectively. Here, we define
\begin{eqnarray}
    \epsilon_{\rm A}(\bm k) &=& -2t_3(\cos k_x+\cos k_y),\\    \epsilon_{\rm B}(\bm k) &=& -2t_3\cos k_x-2t_3^\prime\cos k_y,\\
     \epsilon_{\rm C}(\bm k) &=& -2t_3\cos k_y-2t_3^\prime\cos k_x,\\
     \epsilon_{\rm x}(\bm k) &=& -2t\cos k_x/2,\\
     \epsilon_{\rm y}(\bm k) &=& -2t\cos k_y/2,\\
     \epsilon_{\rm xy}(\bm k) &=& -4t_2\cos k_x/2\cos k_y/2,
\end{eqnarray}
with $(t, t_2, t_3, t_3^\prime) = (1.0, 0.4, 0.15, 0.2)$.
The energy dispersion for the dispersive and usual Lieb lattice models is shown in Figs.~\ref{fig:hk_lieb_all2}(c) and \ref{fig:hk_lieb_all2}(d), respectively. 
Owing to the long-range hopping, the flat band in the original Lieb lattice gets dispersion.
In the dispersive Lieb lattice model, the third-nearest-neighbor hopping of the A sublattice (blue circles in Fig.~\ref{fig:hk_lieb_all2}(a)) is not equivalent to that of the B(C) sublattice (red circles in Fig.~\ref{fig:hk_lieb_all2}(a)). Therefore, three-fold band degeneracy at $M$ point is partially lifted and reduced to two-fold band degeneracy (see Fig.~\ref{fig:hk_lieb_all2}(c)). 

\begin{center}
\end{center}

\begin{center}
\subsection{Non-Kramers band degeneracy at $M$ point}
\end{center}
\
Next, we discuss the non-Kramers band degeneracy in the (dispersive) Lieb lattice at the $M$ point based on the $D_{4h}$ point group symmetry.

\begin{center}
\end{center}

\begin{center}
\subsubsection{Periodic basis of Hamiltonian}
\end{center}
\

Since we adopt the Fourier transform with the internal position of sublattices, the Hamiltonian does not satisfy the Brillouin-zone periodicity, i.e., $H_{\rm 0:d(l)}(\bm k)\neq H_{0:{\rm d(l)}}(\bm k+\bm G)$, with a reciprocal lattice vector $\bm G$. 
Because this basis is not convenient for the symmetry analysis, %\YYS{at $M$ point. To analyze the symmetry constraints,} 
we introduce the Hamiltonian with a periodic basis where the Fourier transform does not include the internal position of sublattices, $\tilde{H}_{\rm 0:d(l)}(\bm k)=V(\bm k, \bm a_{\rm B}, \bm a_{\rm C})H_{\rm 0:d(l)}(\bm k)V^\dagger(\bm k, \bm a_{\rm B}, \bm a_{\rm C})$ with $V(\bm k, \bm a_{\rm B}, \bm a_{\rm C})={\rm diag}(1, e^{i\bm k\cdot \bm a_{\rm B}},e^{i\bm k\cdot \bm a_{\rm C}})$. Here, $\bm a_{\rm B}$ and $\bm a_{\rm C}$ are the internal positions of the sublattices, B and C, respectively.
We set the internal position of the sublattice A as the origin, i.e. $\bm a_{\rm A} = (0,0,0)$.
In this basis, the vector representation of the annihilation operator is written as, $V(\bm k, \bm a_{\rm B}, \bm a_{\rm C}) \, \hat{\bm{c}}(\bm k)$.

\begin{center}
\end{center}

\begin{center}
\subsubsection{Symmetry operation}
\end{center}
\

We consider the symmorphic point group where the point-group element $\hat{g} = \{p_g\}$ does not include any translation operation and its operation on real-space coordinates is given by $\hat{g}\bm r = p_g \bm r$.
The operation of $\hat{g}$ on the Hilbert space is defined by the following relation,
\begin{eqnarray}
  \hat{g}\hat{c}_l(\bm R)\hat{g}^{-1}
    =\hat{c}_{gl}(\bm R^\prime),\\
    p_g \bm R + p_g \bm r_l = \bm R^\prime + \bm r_{gl},   
\end{eqnarray}
 where $gl$ is the transformed sublattice index by the symmetry operation.
 Here and hereafter, we omit the spin index for simplicity.

The wave-vector representation is transformed by the symmetry operation as, 
\begin{eqnarray}
    \hat{g}e^{i\bm k\cdot \bm r_l} \hat{c}_l(\bm k)\hat{g}^{-1} &=&
    \dfrac{1}{\sqrt{N}}e^{i\bm k\cdot \bm r_l}\sum_{\bm R}e^{-i\bm k\cdot(\bm R+\bm r_l)} \hat{c}_{gl}(\bm R^\prime)\notag\\
    &=& \dfrac{1}{\sqrt{N}}\sum_{\bm R}e^{-i p_g\bm k\cdot p_g\bm R} \hat{c}_{gl}(\bm R^\prime)\notag\\
    &=& \dfrac{1}{\sqrt{N}}e^{p_g\bm k\cdot p_g\bm r_l}\sum_{\bm R^\prime}e^{-i p_g\bm k\cdot (\bm R^\prime+\bm r_{gl})} \hat{c}_{gl}(\bm R^\prime)\notag\\
    &=& e^{p_g\bm k\cdot p_g\bm r_l} \hat{c}_{gl}(p_g\bm k).
\end{eqnarray}
Therefore, its vector representation for the dispersive Lieb lattice model can be written by, 
\begin{eqnarray}
    \hat{g}V(\bm k, \bm a_{\rm B}, \bm a_{\rm C})\bm \hat{\bm c}(\bm k)\hat{g}^{-1} &=& V(p_g\bm k, p_g\bm a_{\rm B}, p_g\bm a_{\rm C})D_g\bm \hat{\bm c}(p_g\bm k)\notag\\
    &=& V(p_g\bm k, p_g\bm a_{\rm B}, p_g\bm a_{\rm C})D_gV^\dagger(p_g\bm k, \bm a_{\rm B}, \bm a_{\rm C})V(p_g\bm k, \bm a_{\rm B}, \bm a_{\rm C})\bm \hat{\bm c}(p_g\bm k),\notag\\
\end{eqnarray}
where $[D_{g}]_{ll^\prime} = \delta_{l,gl^\prime}$ is the representation matrix of the symmetry operation $\hat{g}$ with respect to the sublattice degree of freedom.
From this, we obtain the representation matrix for the symmetry operation,
\begin{eqnarray}
U_g(\bm k) = V(p_g\bm k, p_g\bm a_{\rm B}, p_g\bm a_{\rm C}) D_gV^\dagger(p_g\bm k, \bm a_{\rm B}, \bm a_{\rm C}).   
\end{eqnarray}

\begin{center}
\end{center}

\begin{center}
\subsubsection{Symmetry analysis}
\end{center}
\

In the following, we decompose the representation matrix of the symmetry operation at the $M$ point into the irreducible representations of the point group $D_{4h}$. If there is a two-dimensional representation in the decomposition, 
the eigenspectrum at the $M$ point has at least one set of doubly degenerate eigenstates.

For our purpose, it is sufficient to consider $C_{4v}$, a subgroup of $D_{4h}$, since there should be a two-dimensional irreducible representation of $D_{4h}$ when we have that of $C_{4v}$.
The representation matrices at $\bm{k}_M=(\pi,\pi,0)$ are given by
\begin{eqnarray}
    U_{C_4}(\bm k_M)
    &=&
    \left(
        \begin{array}{ccc}
             1&  0&0\\
             0&  0&-1\\
             0&  1&0
        \end{array}
    \right),
    U_{\sigma_{\rm v}}(\bm k_M)
    =
    \left(
        \begin{array}{ccc}
             1&  0&0\\
             0&  1&0\\
             0&  0&-1
        \end{array}
    \right),
    U_{\sigma_{\rm d}}(\bm k_M)
    =
    \left(
        \begin{array}{ccc}
             1&  0&0\\
             0&  0&-1\\
             0&  -1&0
        \end{array}
    \right),\label{eq:repmats}
\end{eqnarray}
and so on.
Here, $g = C_n$ is the $n$ fold rotational symmetry, and 
$\sigma_{v}$ and $\sigma_{d}$ are the mirror reflection symmetry whose mirror planes are rotated from each other by $\pi/4$.

Based on Eq.~\eqref{eq:repmats}, the character table for the $C_{4v}$ point group is summarized in Table~\ref{table:symmetry_analysis}. Characters of the representation matrix for a symmetry operation $\hat{g}$, $\xi_{U_g(\bm k_M)}(g)\equiv{\rm tr}[U_g(\bm{k}_M)]$, are written by $\xi_{U_g(\bm k_M)}(g) = \xi_{A}(g)+\xi_{E}(g)$ with characters of the irreducible representations $\xi_{A}(g)$ and $\xi_{E}(g)$.
As a result, we conclude that the eigenvalues at the $M$ point are generally given by a pair of doubly degenerate eigenstates and a non-degenerate eigenstate.
In the usual Lieb lattice, the condition $\epsilon_{\rm A}(\bm k) = \epsilon_{\rm B(C)}(\bm k)$ leads to the accidental three-fold band degeneracy. 
However, in the dispersive Lieb lattice, the additional third-nearest neighbor hopping lifts the accidental degeneracy while preserving the two-fold band degeneracy protected by the $C_4$ symmetry.

\begin{table}[htbp]
 \caption{The character table of the $C_{4v}$ point group with the representation matrix at the $M$ point.}
 \label{table:symmetry_analysis}
 \centering
  \begin{tabular}{c|ccccc}
& $\bm 1$ & $2C_{4}$ & $C_{2}$ & $2\sigma_{v}$ & $2\sigma_{d}$ \\ \hline
$U_g(\bm k_M)$  & $3$ & $1$ & $-1$ & $1$& $1$ \\
$A_1$ & $1$ & $1$ & $1$ & $1$ & $1$ \\
$A_2$ & $1$ & $1$ & $1$ & $-1$ & $-1$\\
$B_1$ & $-1$ & $1$ & $1$ & $1$ & $-1$ \\
$B_2$ & $1$ & $-1$ & $1$ & $-1$ & $1$ \\
$E$ & $2$ & $0$ & $-2$ & $0$  & $0$ \\
  \end{tabular}
\end{table}

\begin{center}
\end{center}

\begin{center}
\section{Linearized gap equation with RPA}
\end{center}

To study the superconductivity, we solve the following two equations self-consistently:
\begin{eqnarray}
    \lambda\Delta_{ll^\prime}(\bm k) &=& -\dfrac{1}{N\beta}\sum_{\bm k^\prime\omega_n}\sum_{l_1l_2}V_{ll_1,l_2l^\prime}^{\rm t(s)}(\bm k-\bm k^\prime)\mathcal{F}_{l_1l_2}(\bm k^\prime,i\omega_n),\\
    \mathcal{F}_{ll^\prime}(\bm k,i\omega_n) &=& \sum_{l_1l_2}\mathcal{G}_{ll_1}(\bm k,i\omega_n)\Delta_{l_1l_2}(\bm k)\mathcal{G}_{l^\prime l_2}(-\bm k,-i\omega_n).
\end{eqnarray}
Here, $\lambda$ is the eigenvalue of the linearized gap equation and $\Delta_{ll^\prime}(\bm k)$ is the gap function.
The matrix representation of the effective interaction $V_{ll_1,l_2l^\prime}(\bm k)$ for spin-triplet and spin-singlet pairings are obtained by RPA as
\begin{eqnarray}
    V^{\rm t}(\bm k) &=& \Gamma^0\left[-\dfrac{1}{4}\bar{\chi}_{\rm s}(\bm k)-\dfrac{1}{4}\bar{\chi}_{\rm c}(\bm k)\right] \Gamma^0,\\
    V^{\rm s}(\bm k) &=& \Gamma^0\left[\dfrac{3}{4}\bar{\chi}_{\rm s}(\bm k)-\dfrac{1}{4}\bar{\chi}_{\rm c}(\bm k)\right] \Gamma^0 + \Gamma^0,
\end{eqnarray}
respectively.
We ignore the $\Omega_n$-dependence of the effective interaction $V^{\rm t(s)}_{ll_1,l_2l^\prime}(\bm k)$, corresponding to the mean-field approximation.

\begin{center}
\end{center}

\begin{center}
\section{Superconductivity in the dispersive Lieb lattice}
\end{center}

\begin{figure}[htbp]
    \centering
    \includegraphics[width=0.8\linewidth]{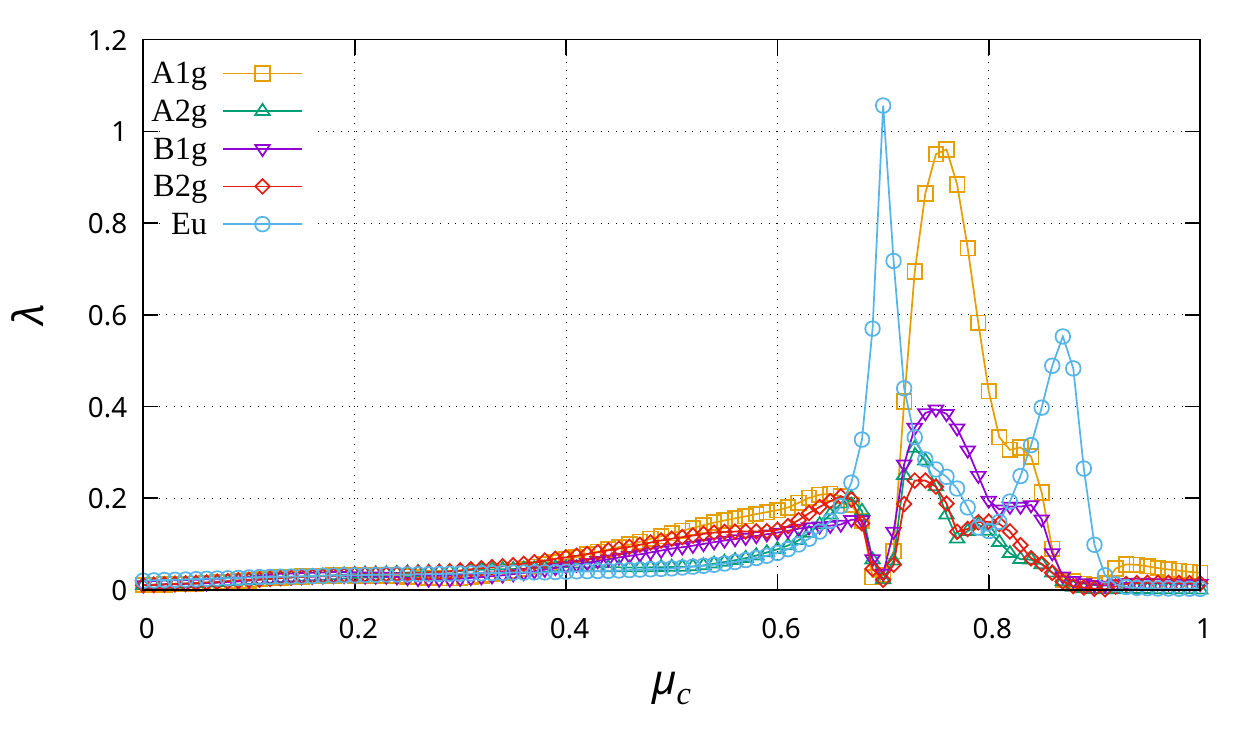}
    \caption{The eigenvalues $\lambda$ of the linearized gap equation for the dispersive Lieb lattice model. We set $T = 0.01$ and $U=0.86$. The orange, green, purple, red, and blue lines correspond to the $A_{1g}, A_{2g}, B_{1g}, B_{2g}$, and $E_u$ representations.}
    \label{fig:lambda_Lieb_all}
\end{figure}

In this section, we show the detailed results of superconductivity in the dispersive Lieb lattice model described by Eq.~\eqref{eq:dispersive_lieb}. The dispersive Lieb lattice belongs to the $D_{4h}$ point group. 
In the presence of $D_{4h}$ point group symmetry, superconducting states are classified into ten irreducible representations.
However, since we consider a purely two-dimensional system with SU(2) symmetry, the $C_2$ rotation is equivalent to the space inversion. Therefore, $A_{1g}, A_{2g}, B_{1g}, B_{2g}$, and $E_u$ representations are allowed while $A_{1u}, A_{2u}, B_{1u}, B_{2u}$, and $E_g$ representations are prohibited.
Here, we set the temperature $T = 0.01$ and show the chemical potential dependence of $\lambda$ for all the allowed irreducible representations in Fig.~\ref{fig:lambda_Lieb_all}. Either the spin-triplet $E_u$ pairing (blue line) or the spin-singlet $A_{1g}$ pairing (orange line) is dominant.

\begin{figure}[htbp]
    \centering
    \includegraphics[width=0.7\linewidth]{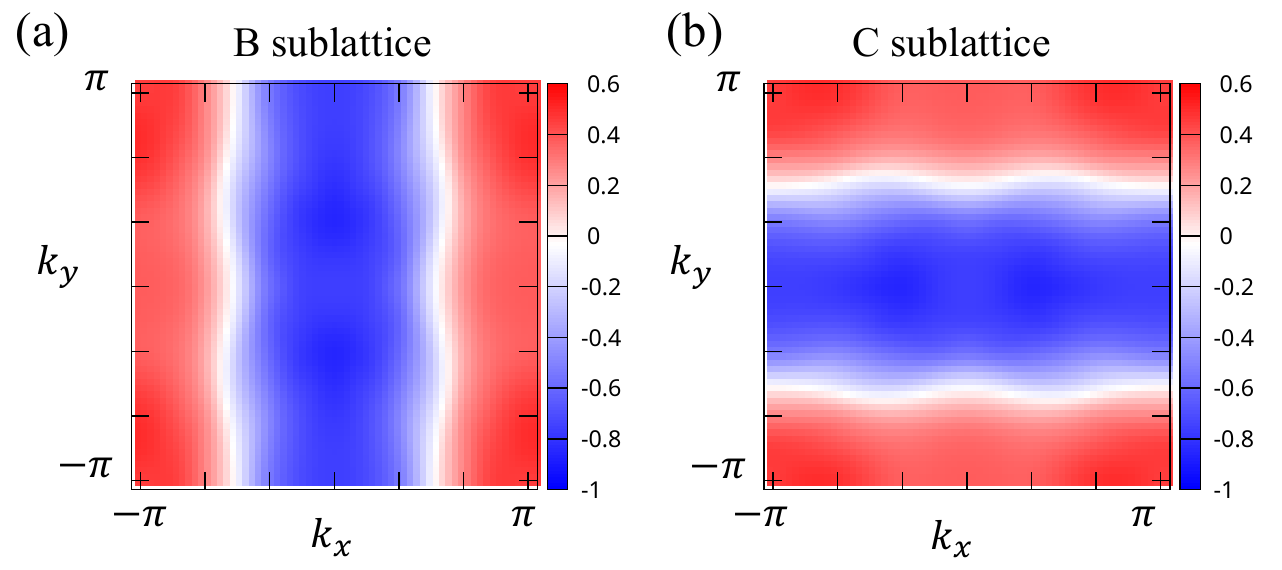}
    \caption{The $\bm k$-dependence of the $A_{1g}$-gap functions at $\mu_{\rm c} = 0.76$ for intra-sublattice pairings on (a) the B sublattice and (b) the C sublattice. As shown in Fig.~\ref{fig:hk_lieb_all2}(a) by red circles, the B (C) sublattice lies on the right (top) of the A sublattice.
    These two sublattices are related to each other by $C_4$ rotation.}
    \label{fig:A1g_Lieb_all.pdf}
\end{figure}

Here, we discuss the dominant $A_{1g}$ and $E_{u}$ superconducting states. 
We show the $\bm k$-dependence of the $A_{1g}$-gap functions at $\mu_{\rm c} = 0.76$ in Fig.~\ref{fig:A1g_Lieb_all.pdf}. 
Figures~\ref{fig:A1g_Lieb_all.pdf}(a) and \ref{fig:A1g_Lieb_all.pdf}(b) show the gap functions for the intra-sublattice pairing on the B sublattice and the C sublattice illustrated in Fig.~\ref{fig:hk_lieb_all2}(a). The other components of the gap function are less dominant than these components. The symmetry of superconductivity corresponds to the extended-$s$-wave superconductivity.

\begin{figure}[htbp]
    \centering
    \includegraphics[width=1.0\linewidth]{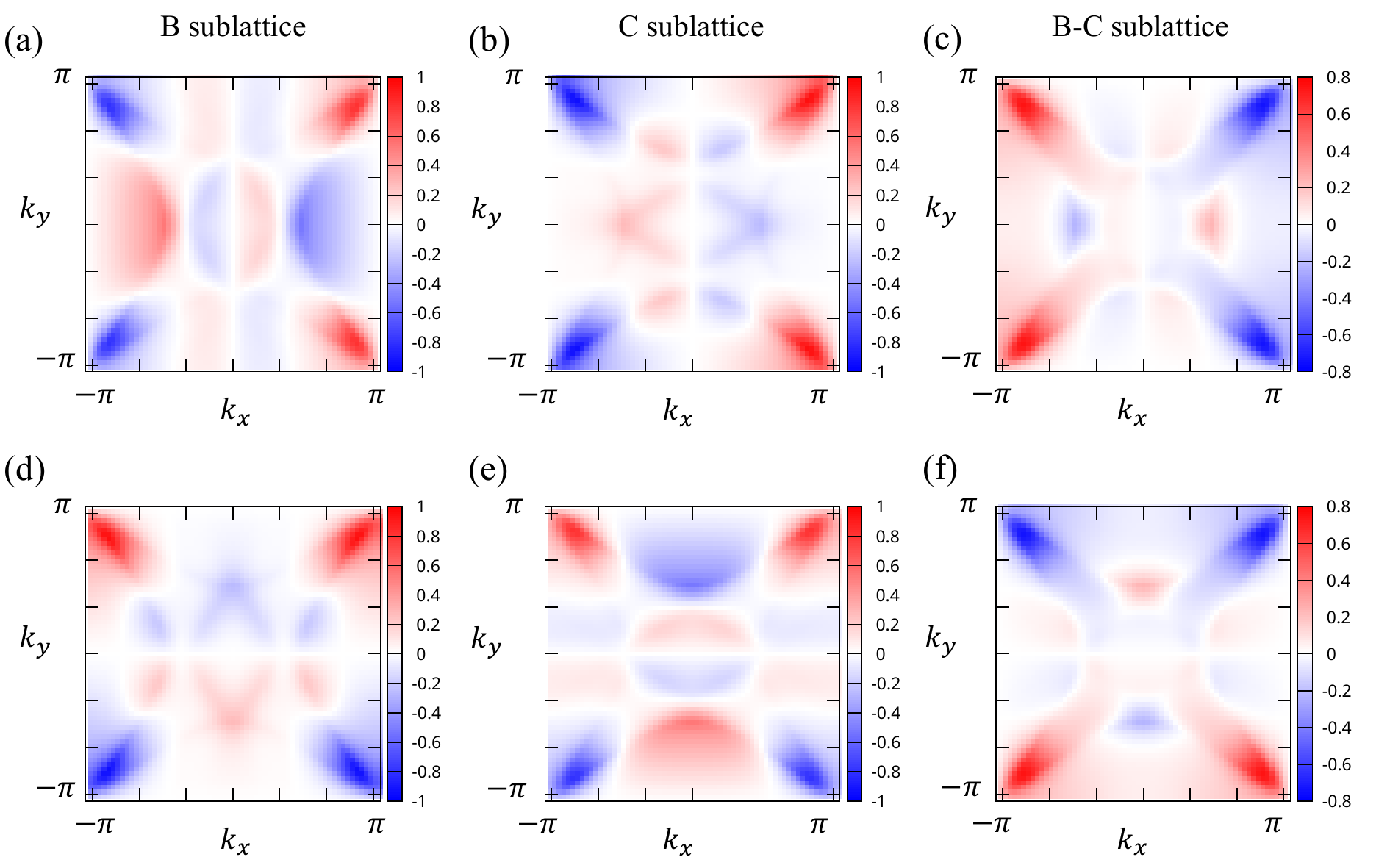}
    \caption{The $\bm k$-dependence of the $E_u$-gap functions at $\mu_{\rm c} = 0.7$. (a)-(c) and (d)-(f) show the $p_x$ and $p_y$ basis of $E_u$ representation, respectively. (a) and (d) [(b) and (e)] are the intra-sublattice pairing component on the B [C] sublattice while (c) and (e) are the inter-sublattice pairing component between the B and C sublattices. }
    \label{fig:Eu_Lieb_all.pdf}
\end{figure}

Figure~\ref{fig:Eu_Lieb_all.pdf} shows the $\bm k$-dependence of the $E_u$-gap functions at $\mu_{\rm c} = 0.7$. In the two-dimensional $E_u$ representation, two independent bases i.e. $p_x$ and $p_y$ are present, corresponding to the degenerate pairing states. Therefore, we show Fig.~\ref{fig:Eu_Lieb_all.pdf}(a)-(c) for the $p_x$ pairing state while Fig.~\ref{fig:Eu_Lieb_all.pdf}(d)-(f) for the $p_y$ state. Panels (a) and (d) [(b) and (e)] are the intra-sublattice pairing component on the B [C] sublattice, while (c) and (f) show the inter-sublattice pairing component between the B and C sublattices. The other components of the gap function are less dominant. In the $E_u$ representation, any linear combination of the two independent bases is allowed. Considering that full-gap superconducting states are thermodynamically stable to maximize the condensation energy, the $p_x+ip_y$-pairing state, namely, the time-reversal-symmetry-broken chiral $p$-wave pairing state may be favored. Some nodes in Fig.~\ref{fig:Eu_Lieb_all.pdf}, such as on the $k_x = 0$ and $k_x = 0$ lines are gapped.
However, the chiral $p$-wave pairing state is degenerate with other $p$-wave pairing states such as $p_x \hat{x} + p_y \hat{y}$ due to the spin degree of freedom of spin-triplet Cooper pairs. Since the degeneracy is protected by the SU(2) symmetry, it is lifted by the SOC.

\begin{center}
\end{center}
\begin{center}
\section{Another example of quantum-geometry-induced ferromagnetic fluctuation : Raghu's model}
\end{center}

\begin{figure}[htbp]
    \centering
    \includegraphics[width=0.8\linewidth]{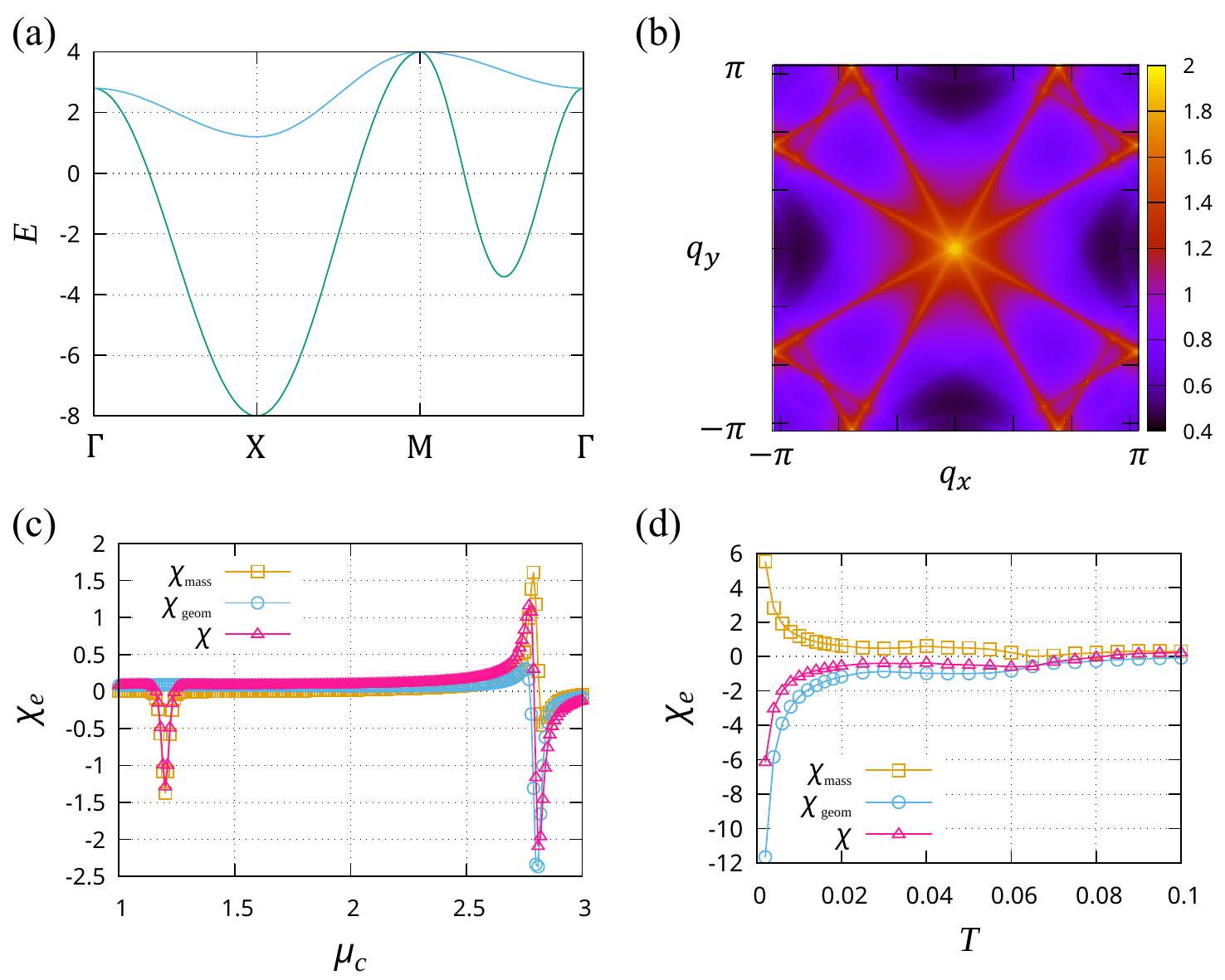}
    \caption{Results in the Raghu's model. (a) The band dispersion, (b) the bare spin susceptibility $\chi_{\rm s}^{0}(\bm q)$  for $(\mu_{\rm c}, T) = (2.8, 0.002)$, (c) the chemical potential dependence of the generalized electric susceptibility $\chi_{\rm e}^{0:xx}$ at $T=0.01$, and (d) the temperature dependence of $\chi_{\rm e}^{0:xx}$ at $\mu_{\rm c} = 2.8$.
    The geometric term and effective-mass term are also shown in (c) and (d).}
    \label{fig:Raghu_all.pdf}
\end{figure}

To show another example of quantum-geometry-induced ferromagnetic fluctuation, we consider Raghu's model~\cite{Sraghu2008minimal} for iron-based superconductors. Using the Pauli matrix and the unit matrix for the orbital space $\rho_{\mu}$ and $\rho_0$, the Hamiltonian of the Raghu's model is given by,
\begin{eqnarray}
    H_{0:{\rm r}}(\bm k) = (h_0(\bm k)-\mu_{\rm c})\rho_0 + h_{\rm xy}(\bm k)\rho_x + h_{\rm z}(\bm k)\rho_z.
\end{eqnarray}
Here, we define,
\begin{eqnarray}
    h_0(\bm k) &=& -(t_1+t_2)(\cos k_x+\cos k_y) -4t_3\cos k_x\cos k_y,\\
    h_{z}(\bm k) &=& -(t_1-t_2)(\cos k_x-\cos k_y),\\
    h_{xy}(\bm k) &=& -4t_4\sin k_x \sin k_y,
\end{eqnarray}
with $(t_1, t_2, t_3, t_4) = (-1.0, 1.3, -0.85, -0.85)$.

The band dispersion of the Raghu's model is shown in Fig.~\ref{fig:Raghu_all.pdf}(a). We see the band degeneracy at $\Gamma$ and $M$ points.
The generalized electric susceptibility in this model is shown Figs.~\ref{fig:Raghu_all.pdf}(c) and \ref{fig:Raghu_all.pdf}(d), where the total susceptibility $\chi_{\rm e}^{0:xx}$, the quantum geometric term $\chi_{\rm e:geom}^{0:xx}$, and the 
effective-mass term $\chi_{\rm e:mass}^{0:xx}$ are plotted. 
In Fig.~\ref{fig:Raghu_all.pdf}(c) showing the chemical potential dependence, we find the negative peak of $\chi_{\rm e:geom}^{0:xx}$ near $\mu_{\rm c} = 2.8$, where the band-degenerate point lies on the Fermi surface. Furthermore, the geometric term $\chi_{\rm e:geom}^{0:xx}$ is negatively enhanced as the temperature decreases (see Fig.~\ref{fig:Raghu_all.pdf}(d) for the temperature dependence). These behaviors are similar to the dispersive Lieb lattice model discussed in the main text and indicate the quantum-geometry-induced ferromagnetic fluctuation. The bare spin susceptibility $\chi_{\rm s}^{0}({\bm q})$ with the peak at ${\bm q}=0$ is shown in Fig.~\ref{fig:Raghu_all.pdf}(b), and the ferromagnetic fluctuation is confirmed. Thus, we conclude that the quantum geometry induces ferromagnetic fluctuation due to non-Kramers band degeneracy also in the Raghu's model.


\begin{thebibliography}{76}
\bibitem [{qch()}]{qchicomment}%
  \BibitemOpen
  \href@noop {} {}\bibinfo {note} {Since the relationship $\chi_{\rm s(c)}(\bm
  q, i\Omega_n) = \chi_{\rm s (c)}(-\bm q, -i\Omega_n)^*$ is always satisfied,
  the odd-order derivative of $\chi(\bm q,0) (= \chi(\bm q,0)^*)$ with respect
  to $\bm q$ vanishes at $\bm{q}=0$.}\BibitemShut {Stop}%
\end{thebibliography}
\end{document}